\documentstyle[11pt,epsf]{article}
\def\xx{\vrule height 0.3em depth 0.2em width 0.3em}
\def\ra{\rightarrow}
\def\longra{\longrightarrow}

\def\fl{\forall}
\def\ify{\infty}
\def\lgl{\langle}
\def\op{\oplus}
\def\ot{\otimes}
\def\ov{\overline}
\def\rgl{\rangle}
\def\ts{\times}
\def\wt{\widetilde}

\def\a{\alpha}
\def\b{\beta}
\def\d{\delta}
\def\g{\gamma}
\def\s{\sigma}
\def\t{\theta}
\def\ve{\varepsilon}
\def\vp{\varphi}

\def\D{\Delta}
\def\G{\Gamma}

\font\tenbb=msbm10
\font\sevenbb=msbm7
\font\fivebb=msbm5
\newfam\bbfam
\textfont\bbfam=\tenbb \scriptfont\bbfam=\sevenbb
\scriptscriptfont\bbfam=\fivebb
\def\bb{\fam\bbfam}

\def\Cb{{\bb C}}
\def\Nb{{\bb N}}
\def\Rb{{\bb R}}

\def\Hc{{\cal H}}

\catcode`\@=11
\def\displaylinesno #1{\displ@y\halign{
\hbox to\displaywidth{$\@lign\hfil\displaystyle##\hfil$}&
\llap{$##$}\crcr#1\crcr}}

\def\ldisplaylinesno #1{\displ@y\halign{
\hbox to\displaywidth{$\@lign\hfil\displaystyle##\hfil$}&
\kern-\displaywidth\rlap{$##$}
\tabskip\displaywidth\crcr#1\crcr}}
\catcode`\@=12

\def\semi{\mathop{>\!\!\!\triangleleft}}

\def\build#1_#2^#3{\mathrel{
\mathop{\kern 0pt#1}\limits_{#2}^{#3}}}

\def\ot{\otimes}
\def\v{\;\raisebox{-1.5mm}{\epsfysize=6mm\epsfbox{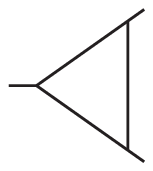}}\;}
\def\vlv{\;\raisebox{-1.5mm}{\epsfysize=6mm\epsfbox{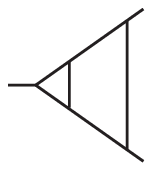}}\;}
\def\vuv{\;\raisebox{-1.5mm}{\epsfysize=6mm\epsfbox{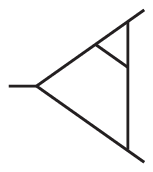}}\;}
\def\vdv{\;\raisebox{-1.5mm}{\epsfysize=6mm\epsfbox{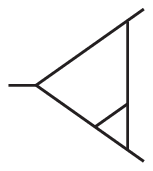}}\;}
\def\vup{\;\raisebox{-1.5mm}{\epsfysize=6mm\epsfbox{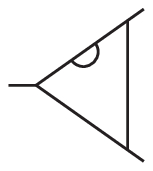}}\;}
\def\vdp{\;\raisebox{-1.5mm}{\epsfysize=6mm\epsfbox{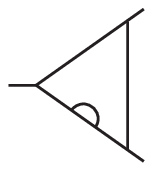}}\;}
\def\vrp{\;\raisebox{-1.5mm}{\epsfysize=6mm\epsfbox{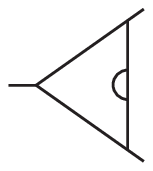}}\;}
\def\w{\;\raisebox{-1.5mm}{\epsfysize=6mm\epsfbox{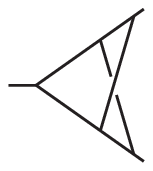}}\;}
\def\vlvlv{\;\raisebox{-1.5mm}{\epsfysize=6mm\epsfbox{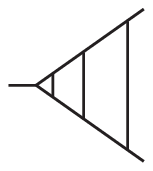}}\;}
\def\vlvuv{\;\raisebox{-1.5mm}{\epsfysize=6mm\epsfbox{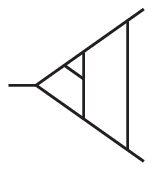}}\;}
\def\vlvdv{\;\raisebox{-1.5mm}{\epsfysize=6mm\epsfbox{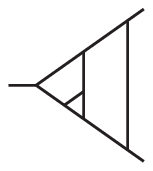}}\;}
\def\vlvup{\;\raisebox{-1.5mm}{\epsfysize=6mm\epsfbox{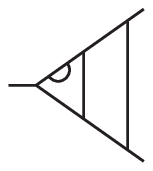}}\;}
\def\vlvdp{\;\raisebox{-1.5mm}{\epsfysize=6mm\epsfbox{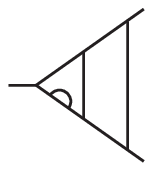}}\;}
\def\vlvrp{\;\raisebox{-1.5mm}{\epsfysize=6mm\epsfbox{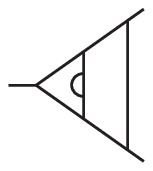}}\;}
\def\vlw{\;\raisebox{-1.5mm}{\epsfysize=6mm\epsfbox{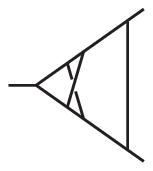}}\;}
\def\vuvlv{\;\raisebox{-1.5mm}{\epsfysize=6mm\epsfbox{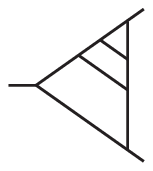}}\;}
\def\vuvuv{\;\raisebox{-1.5mm}{\epsfysize=6mm\epsfbox{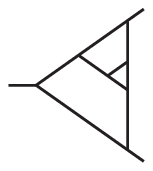}}\;}
\def\vuvdv{\;\raisebox{-1.5mm}{\epsfysize=6mm\epsfbox{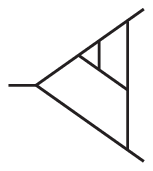}}\;}
\def\vuvup{\;\raisebox{-1.5mm}{\epsfysize=6mm\epsfbox{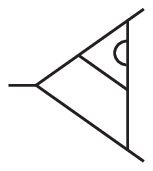}}\;}
\def\vuvdp{\;\raisebox{-1.5mm}{\epsfysize=6mm\epsfbox{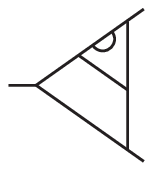}}\;}
\def\vuvrp{\;\raisebox{-1.5mm}{\epsfysize=6mm\epsfbox{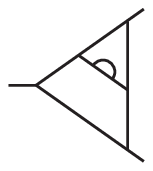}}\;}
\def\vuw{\;\raisebox{-1.5mm}{\epsfysize=6mm\epsfbox{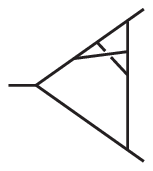}}\;}
\def\vdvlv{\;\raisebox{-1.5mm}{\epsfysize=6mm\epsfbox{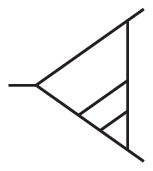}}\;}
\def\vdvuv{\;\raisebox{-1.5mm}{\epsfysize=6mm\epsfbox{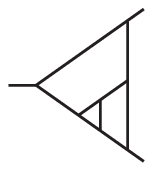}}\;}
\def\vdvdv{\;\raisebox{-1.5mm}{\epsfysize=6mm\epsfbox{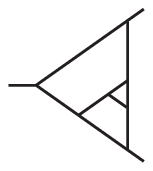}}\;}
\def\vdvup{\;\raisebox{-1.5mm}{\epsfysize=6mm\epsfbox{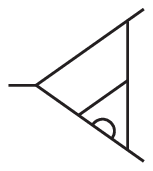}}\;}
\def\vdvdp{\;\raisebox{-1.5mm}{\epsfysize=6mm\epsfbox{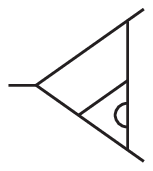}}\;}
\def\vdvrp{\;\raisebox{-1.5mm}{\epsfysize=6mm\epsfbox{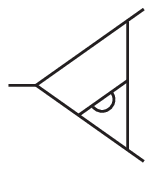}}\;}
\def\vdw{\;\raisebox{-1.5mm}{\epsfysize=6mm\epsfbox{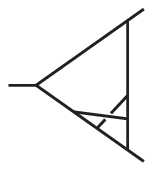}}\;}
\def\vluvv{\;\raisebox{-1.5mm}{\epsfysize=6mm\epsfbox{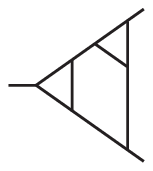}}\;}
\def\vudvv{\;\raisebox{-1.5mm}{\epsfysize=6mm\epsfbox{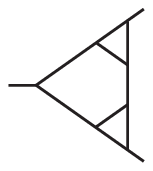}}\;}
\def\vldvv{\;\raisebox{-1.5mm}{\epsfysize=6mm\epsfbox{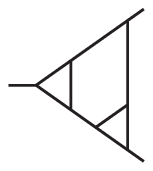}}\;}
\def\vldvp{\;\raisebox{-1.5mm}{\epsfysize=6mm\epsfbox{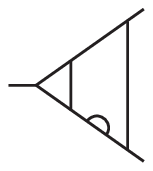}}\;}
\def\vlrvp{\;\raisebox{-1.5mm}{\epsfysize=6mm\epsfbox{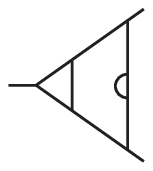}}\;}
\def\vluvp{\;\raisebox{-1.5mm}{\epsfysize=6mm\epsfbox{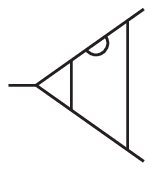}}\;}
\def\vdrvp{\;\raisebox{-1.5mm}{\epsfysize=6mm\epsfbox{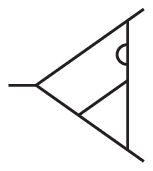}}\;}
\def\vduvp{\;\raisebox{-1.5mm}{\epsfysize=6mm\epsfbox{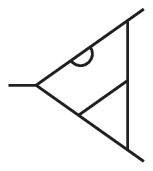}}\;}
\def\vddvp{\;\raisebox{-1.5mm}{\epsfysize=6mm\epsfbox{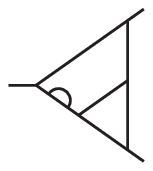}}\;}
\def\vuuvp{\;\raisebox{-1.5mm}{\epsfysize=6mm\epsfbox{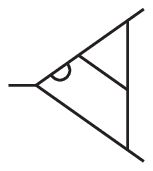}}\;}
\def\vudvp{\;\raisebox{-1.5mm}{\epsfysize=6mm\epsfbox{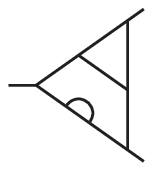}}\;}
\def\vurvp{\;\raisebox{-1.5mm}{\epsfysize=6mm\epsfbox{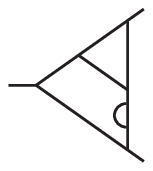}}\;}
\def\p{\;\raisebox{-1.5mm}{\epsfysize=6mm\epsfbox{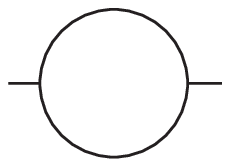}}\;}
\def\pv{\;\raisebox{-1.5mm}{\epsfysize=6mm\epsfbox{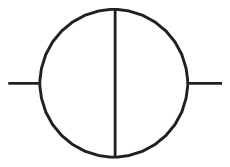}}\;}
\def\pdp{\;\raisebox{-1.5mm}{\epsfysize=6mm\epsfbox{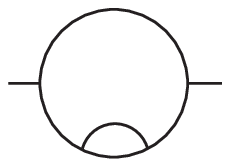}}\;}
\def\pdpdp{\;\raisebox{-1.5mm}{\epsfysize=6mm\epsfbox{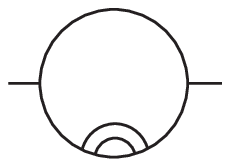}}\;}
\def\pdpv{\;\raisebox{-1.5mm}{\epsfysize=6mm\epsfbox{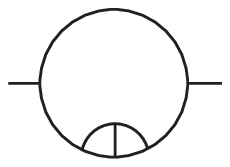}}\;}
\def\pddpp{\;\raisebox{-1.5mm}{\epsfysize=6mm\epsfbox{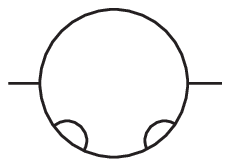}}\;}
\def\pdupp{\;\raisebox{-1.5mm}{\epsfysize=6mm\epsfbox{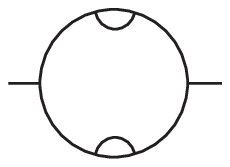}}\;}
\def\pw{\;\raisebox{-1.5mm}{\epsfysize=6mm\epsfbox{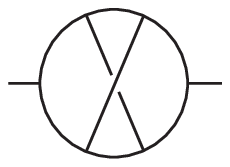}}\;}
\def\pvmp{\;\raisebox{-1.5mm}{\epsfysize=6mm\epsfbox{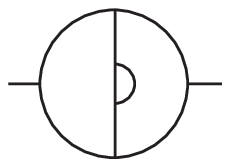}}\;}
\def\pvlp{\;\raisebox{-1.5mm}{\epsfysize=6mm\epsfbox{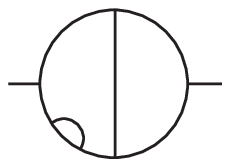}}\;}
\def\pvrp{\;\raisebox{-1.5mm}{\epsfysize=6mm\epsfbox{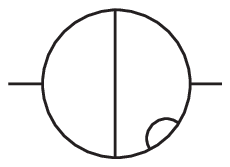}}\;}
\def\plrvv{\;\raisebox{-1.5mm}{\epsfysize=6mm\epsfbox{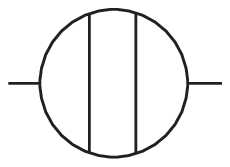}}\;}
\def\pvuv{\;\raisebox{-1.5mm}{\epsfysize=6mm\epsfbox{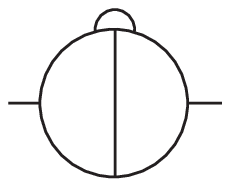}}\;}
\def\wlv{\;\raisebox{-1.5mm}{\epsfysize=6mm\epsfbox{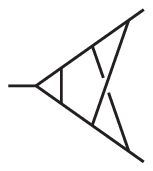}}\;}
\def\wulv{\;\raisebox{-1.5mm}{\epsfysize=6mm\epsfbox{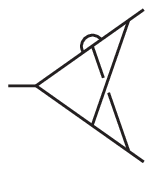}}\;}
\def\wurv{\;\raisebox{-1.5mm}{\epsfysize=6mm\epsfbox{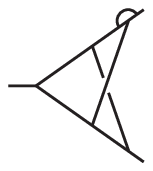}}\;}
\def\wdrv{\;\raisebox{-1.5mm}{\epsfysize=6mm\epsfbox{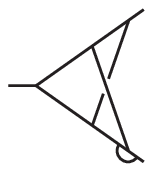}}\;}
\def\wulp{\;\raisebox{-1.5mm}{\epsfysize=6mm\epsfbox{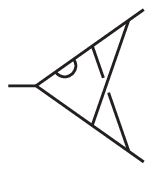}}\;}
\def\wurp{\;\raisebox{-1.5mm}{\epsfysize=6mm\epsfbox{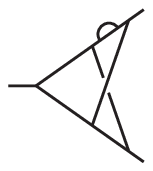}}\;}
\def\wdlp{\;\raisebox{-1.5mm}{\epsfysize=6mm\epsfbox{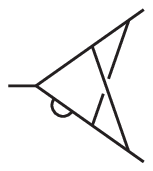}}\;}
\def\wdrp{\;\raisebox{-1.5mm}{\epsfysize=6mm\epsfbox{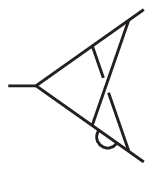}}\;}
\def\wmrp{\;\raisebox{-1.5mm}{\epsfysize=6mm\epsfbox{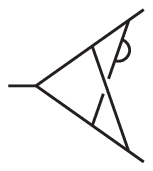}}\;}
\def\wmlp{\;\raisebox{-1.5mm}{\epsfysize=6mm\epsfbox{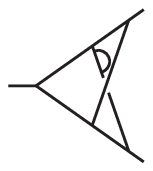}}\;}

\def\vupv{\;\raisebox{-1.5mm}{\epsfysize=6mm\epsfbox{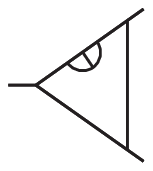}}\;}
\def\vdpv{\;\raisebox{-1.5mm}{\epsfysize=6mm\epsfbox{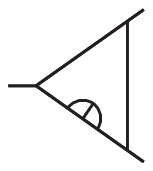}}\;}
\def\vrpv{\;\raisebox{-1.5mm}{\epsfysize=6mm\epsfbox{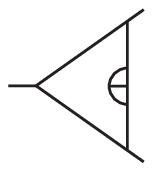}}\;}
\def\vupdp{\;\raisebox{-1.5mm}{\epsfysize=6mm\epsfbox{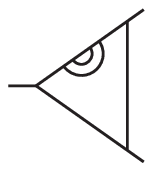}}\;}
\def\vdpdp{\;\raisebox{-1.5mm}{\epsfysize=6mm\epsfbox{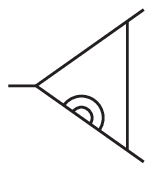}}\;}
\def\vrpdp{\;\raisebox{-1.5mm}{\epsfysize=6mm\epsfbox{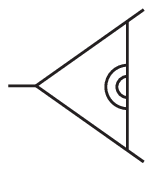}}\;}
\def\vuupp{\;\raisebox{-1.5mm}{\epsfysize=6mm\epsfbox{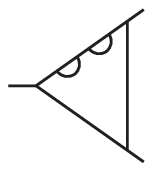}}\;}
\def\vrrpp{\;\raisebox{-1.5mm}{\epsfysize=6mm\epsfbox{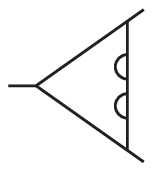}}\;}
\def\vddpp{\;\raisebox{-1.5mm}{\epsfysize=6mm\epsfbox{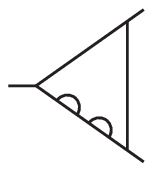}}\;}
\def\vurpp{\;\raisebox{-1.5mm}{\epsfysize=6mm\epsfbox{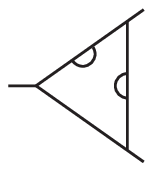}}\;}
\def\vdrpp{\;\raisebox{-1.5mm}{\epsfysize=6mm\epsfbox{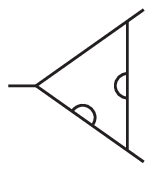}}\;}
\def\vudpp{\;\raisebox{-1.5mm}{\epsfysize=6mm\epsfbox{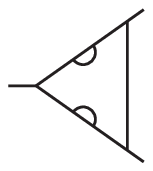}}\;}

\begin{document}
\begin{center}
{\large {\bf  Renormalization in quantum field theory and the
Riemann-Hilbert problem II: the $\beta$-function, diffeomorphisms
and the renormalization group} \footnote{IHES/M/00/22,
hep-th/0003188; connes@ihes.fr, kreimer@ihes.fr}\\[1cm]} {\bf
Alain Connes and Dirk Kreimer\\[5mm]} {\small Institut des Hautes
\'Etudes Scientifiques, March 2000\\[2cm]}
\end{center}


{\footnotesize \noindent {\bf Abstract } } {\em We showed in part
I that the Hopf algebra ${\cal H}$ of Feynman graphs in a given
QFT is the algebra of coordinates on a complex infinite
dimensional Lie group $G$ and that the renormalized theory is
obtained from the unrenormalized one by evaluating at $\ve=0$ the
holomorphic part $\gamma_+(\ve)$ of the Riemann--Hilbert
decomposition $\gamma_-(\ve)^{-1}\gamma_+(\ve)$ of the loop
$\gamma(\ve)\in G$ provided by dimensional regularization. We show
in this paper that the group $G$ acts naturally on the complex
space $X$ of dimensionless coupling constants of the theory. More
precisely, the formula $g_0=gZ_1Z_3^{-3/2}$ for the effective
coupling constant, when viewed as a formal power series, does
define a Hopf algebra homomorphism between the Hopf algebra of
coordinates on the group of formal diffeomorphisms to the Hopf
algebra ${\cal H}$. This allows first of all to read off directly,
without using the group $G$, the bare coupling constant and the
renormalized one from the Riemann--Hilbert decomposition of the
unrenormalized effective coupling constant viewed as a loop of
formal diffeomorphisms. This shows that renormalization is
intimately related with the theory of non-linear complex bundles
on the Riemann sphere of the dimensional regularization parameter
$\ve$. It also allows to lift both the renormalization group and
the $\beta$-function as the asymptotic scaling in the group $G$.
This exploits the full power of the Riemann--Hilbert decomposition
together with the invariance of $\gamma_-(\ve)$ under a change of
unit of mass. This not only gives a conceptual proof of the
existence of the renormalization group but also delivers a
scattering formula in the group $G$ for the full higher pole
structure of minimal subtracted counterterms in terms of the
residue.}
\newpage

\section{Introduction}\label{secI}
We showed in part~I of this paper \cite{I}
that perturbative
renormalization is a special case of a general mathematical
procedure of extraction of finite values based on the
Riemann-Hilbert problem. More specifically we associated to any
given renormalizable quantum field theory an (infinite
dimensional) complex Lie group $G$. We then showed that passing
from the unrenormalized theory to the renormalized one was exactly
the replacement of the loop $d \ra \g (d) \in G$ of elements of
$G$ obtained from dimensional regularization (for $d \ne D =$
dimension of space-time) by the value $\g_+ (D)$ of its Birkhoff
decomposition, $\g (d) = \g_- (d)^{-1} \, \g_+ (d)$.

The original loop $d \ra \g (d)$ not only depends upon the
parameters of the theory but also on the additional ``unit of
mass'' $\mu$ required by dimensional analysis. We shall show in
this paper that the mathematical concepts developped in part~I
provide very powerful tools to lift the usual concepts of the
$\b$-function and renormalization group from the space of coupling
constants of the theory to the complex Lie group $G$.

We first observe, taking $\vp_6^3$ as an illustrative example to
fix ideas and notations, that even though the loop $\g (d)$ does
depend on the additional parameter $\mu$, $$ \mu \ra \g_{\mu} (d)
\, , \leqno (1) $$ the negative part $\g_{\mu^-}$ in the Birkhoff
decomposition, $$ \g_{\mu} (d) = \g_{\mu^-} (d)^{-1} \, \g_{\mu^+}
(d) \leqno (2) $$ is actually independent of $\mu$, $$
\frac{\partial}{\partial \mu} \, \g_{\mu^-} (d) = 0 \, . \leqno
(3) $$ This is a restatement of a well known fact and follows
immediately from dimensional analysis. Moreover, by construction,
the Lie group $G$ turns out to be graded, with grading, $$ \t_t
\in {\rm Aut} \, G \ , \quad t \in \Rb \, , \leqno (4) $$
inherited from the grading of the Hopf algebra $\Hc$ of Feynman
graphs given by the loop number, $$ L (\G) = \hbox{loop number of}
\ \G \leqno (5) $$ for any 1PI graph $\G$.

The straightforward equality, $$ \g_{e^t \mu} (d) = \t_{t \ve}
(\g_{\mu} (d)) \qquad \fl \, t \in \Rb \, , \ \ve = D-d \leqno (6)
$$ shows that the loops $\g_{\mu}$ associated to the
unrenormalized theory satisfy the striking property that the
negative part of their Birkhoff decomposition is unaltered by the
operation, $$ \g (\ve) \ra \t_{t\ve} (\g (\ve)) \, . \leqno (7) $$
In other words, if we replace $\g (\ve) $ by $\t_{t\ve} (\g
(\ve))$ we do not change the negative part of its Birkhoff
decomposition. We settled now for the variable, $$ \ve = D - d \in
\Cb \backslash \{ 0 \} \, . \leqno (8) $$ Our first result
(section~\ref{secII}) is a complete characterization of the loops $\g (\ve)
\in G$ fulfilling the above striking invariance. This
characterization only involves the negative part $\g_- (\ve)$ of
their Birkhoff decomposition which by hypothesis fulfills, $$ \g_-
(\ve) \, \t_{t \ve} (\g_- (\ve)^{-1}) \ \hbox{is convergent for} \
\ve \ra 0 \, . \leqno (9) $$ It is easy to see that the limit of
(9) for $\ve \ra 0$ defines a one parameter subgroup, $$ F_t \in G
\, , \ t \in \Rb \leqno (10) $$ and that the generator $\b =
\left( \frac{\partial}{\partial t} \, F_t \right)_{t=0}$ of this
one parameter group is related to the {\it residue} of $\g$ $$
\build{\rm Res}_{\ve = 0}^{} \g = - \left( \frac{\partial}{\partial
u} \, \g_- \left( \frac{1}{u} \right) \right)_{u=0} \leqno (11) $$
by the simple equation, $$ \b = Y \, {\rm Res} \, \g \, , \leqno
(12) $$ where $Y = \left( \frac{\partial}{\partial t} \, \t_t
\right)_{t=0}$ is the grading.

This is straightforward but our result is the following formula
(14) which gives $\g_- (\ve)$ in closed form as a function of
$\b$. We shall for convenience introduce an additional generator
in the Lie algebra of $G$ (i.e.~primitive elements of $\Hc^*$)
such that, $$ [Z_0 , X] = Y(X) \qquad \fl \, X \in \hbox{Lie} \ G
\, . \leqno (13) $$ The scattering formula for $\g_- (\ve)$ is
then, $$ \g_- (\ve) = \lim_{t \ra \ify} e^{-t \left(
\frac{\b}{\ve} + Z_0 \right)} \, e^{t Z_0} \, . \leqno (14) $$
Both factors in the right hand side belong to the semi-direct
product, $$ \wt G = G \, \semi_{\t} \, \Rb \leqno (15) $$ of the
group $G$ by the grading, but of course the ratio (14) belongs to
the group $G$.

This shows (section~\ref{secIII}) that the higher pole structure of the
divergences is uniquely determined by the residue and gives a
strong form of the t'Hooft relations, which will come as an
immediate corollary.

In section~\ref{sec4} we show, specializing to the massless case, that the
formula for the bare coupling constant, $$ g_0 = g \, Z_1 \,
Z_3^{-3/2} \leqno (16) $$ where both $g \, Z_1 =g + \d g$ and the
field strength renormalization constant $Z_3$ are thought of as
power series (in $g$) of elements of the Hopf algebra $\Hc$, does
define a Hopf algebra homomorphism, $$ \Hc_{CM} \build
\longra_{}^{g_0} \Hc \, , \leqno (17) $$ from the Hopf algebra
$\Hc_{CM}$ of coordinates on the group of formal diffeomorphisms
of $\Cb$ such that, $$ \vp (0) = 0 \, , \ \vp' (0) = {\rm id}
\leqno (18) $$ to the Hopf algebra $\Hc$ of the massless theory.
We had already constructed in \cite{CK} a Hopf algebra
homomorphism from $\Hc_{\rm CM}$ to the Hopf algebra of rooted
trees, but the physical significance of this construction was
unclear.

The homomorphism (17) is quite different in that for instance the
transposed group homomorphism, $$ G \build \longra_{}^{\rho} {\rm
Diff} (\Cb) \leqno (19) $$ lands in the subgroup of {\it odd}
diffeomorphisms, $$ \vp (-z) = -\vp (z) \qquad \fl \, z \, .
\leqno (20) $$ Moreover its physical significance will be
transparent. We shall show in particular that the image by $\rho$
of $\b = Y \, {\rm Res} \, \g$ is the usual $\b$-function of the
coupling constant $g$.

We discovered the homomorphism (17) by lengthy concrete
computations. We have chosen to include them in an appendix
besides our conceptual proof given in section~4. The main reason
for this choice is that the explicit computation allows to
validate the concrete ways of handling the coproduct,
coassociativity, symmetry factors$\ldots$ that underly the theory.

As a corollary of the construction of $\rho$ we get an {\it
action} by (formal) diffeomorphisms of the group $G$ on the space
$X$ of (dimensionless) coupling constants of the theory. We can
then in particular formulate the Birkhoff decomposition {\it
directly} in the group, $$ {\rm Diff} \, (X) \leqno (21) $$ of
formal diffeomorphisms of the space of coupling constants.

The unrenormalized theory delivers a loop $$ \d (\ve) \in {\rm
Diff} \, (X) \, , \ \ve \ne 0 \leqno (22) $$ whose value at $\ve$
is simply the unrenormalized effective coupling constant.

The Birkhoff decomposition, $$ \d (\ve) = \, \d_+
(\ve) \, \d_- (\ve)^{-1} \leqno (23) $$ of this loop gives directly then, $$ \d_-
(\ve) = \hbox{ bare coupling constant} \leqno (24) $$ and, $$ \d_+
(D) = \hbox{renormalized effective coupling constant.} \leqno (25)
$$ This result is now, in its statement, no longer depending upon
our group $G$ or the Hopf algebra $\Hc$. But of course the proof
makes heavy use of the above ingredients.

Now the Birkhoff decomposition of a loop, $$ \d (\ve) \in {\rm
Diff} \, (X) \, , \leqno (26) $$ admits a beautiful geometric
interpretation. If we let $X$ be a complex manifold and pass from
formal diffeomorphisms to actual ones, the data (26) is the initial
data to perform, by the clutching operation, the construction of a
complex bundle, $$ P = (S^+ \ts X) \, \cup_{\d} (S^- \ts X) \leqno
(27) $$ over the sphere $S = P_1 (\Cb) = S^+ \cup S^-$, and with
fiber $X$, $$ X \longra P \build \longra_{}^{\pi} S \, . \leqno
(28) $$ The meaning of the Birkhoff decomposition (23), $$ \d (\ve) =
 \d_+ \, (\ve) \, \d_- (\ve)^{-1}  $$ is then exactly
captured by an isomorphism of the bundle $P$ with the trivial
bundle, $$ S \ts X \, . \leqno (29) $$

\section{Asymptotic scaling in graded complex Lie groups}\label{secII}
We shall first prove the formula (14) of the introduction in the
general context of graded Hopf algebras and then apply it to the
Birkhoff decomposition of the loop associated in part~I to the
unrenormalized theory.

We let $\Hc$ be a connected commutative graded Hopf algebra
(connected means that $\Hc^{(0)} = \Cb$) and let $\t_t$, $t \in
\Rb$ be the one parameter group of automorphisms of $\Hc$
associated with the grading so that for $x \in \Hc$ of degree $n$,
$$ \t_t (x) = e^{tn} \, x \qquad \fl \, t \in \Rb \, . \leqno (1)
$$ By construction $\t_t$ is a Hopf algebra automorphism, $$ \t_t
\in {\rm Aut} (\Hc) \, . \leqno (2) $$ We also let $Y = \left(
\frac{\partial}{\partial t} \, \t_t \right)_{t=0}$ be the
generator which is a derivation of $\Hc$.

We let $G$ be the group of characters of $\Hc$, $$ \vp : \Hc \ra
\Cb \leqno (3) $$ i.e. of homomorphisms from the algebra $\Hc$ to
$\Cb$. The product in $G$ is given by, $$ (\vp_1 \, \vp_2) (x) =
\lgl \vp_1 \ot \vp_2 \, , \, \D x \rgl \leqno (4) $$  where $\D$ is
the coproduct in $\Hc$. The augmentation $\ov e$ of $\Hc$ is the
unit of $G$ and the inverse of $\vp \in G$ is given by, $$ \lgl
\vp^{-1} , x \rgl = \lgl \vp , S x \rgl \leqno (5) $$ where $S$ is
the antipode in $\Hc$.

We let $L$ be the Lie algebra of derivations, $$ \d : \Hc \ra \Cb
\, , \leqno (6) $$ i.e. of linear maps on $\Hc$ such that $$ \d
(xy) = \d (x) \, \ov e (y) + \ov e (x) \, \d (y) \qquad \fl \, x,y
\in \Hc \, . \leqno (7) $$ Even if $\Hc$ is of finite type so that
$\Hc^{(n)}$ is finite dimensional for any $n \in \Nb$ there are
more elements in $L$ than in the Lie algebra $P$ of primitive
elements, $$ \D Z = Z \ot 1 + 1 \ot Z \leqno (8) $$ in the graded
dual Hopf algebra $\Hc_{\rm gr}^*$ of $\Hc$. But one passes from
$P$ to $L$ by completion relative to the $I$-adic topology, $I$
being the augmentation ideal of $\Hc_{\rm gr}^*$.

The linear dual $\Hc^*$ is in general an algebra (with product
given by (4)) but not a Hopf algebra since the coproduct is not
necessarily well defined. It is however well defined for
characters $\vp$ or derivations $\d$ which satisfy respectively
$\D \vp = \vp \ot \vp$, and $\D \d = \d \ot 1 + 1 \ot \d$.

For $\d \in L$ the expression, $$ \vp = \exp \, \d \leqno (9) $$
makes sense in the algebra $\Hc^*$ since when evaluated on $x \in
\Hc$ one has $\lgl x , \d^n \rgl = 0$ for $n$ large enough (since
$\lgl x , \d^n \rgl = \lgl \D^{(n-1)} x , \d \ot \cdots \ot \d
\rgl$ vanishes for $n > \deg x$). Moreover $\vp$ is a group-like
element of $\Hc^*$, i.e. a character of $\Hc$. Thus $\vp \in G$.

The one parameter group $\t_t \in {\rm Aut} \, (\Hc)$ acts by
automorphisms on the group $G$, $$ \lgl \t_t (\vp) , x \rgl = \lgl
\vp , \t_t (x) \rgl \qquad \fl \, x \in \Hc \leqno (10) $$ and the
derivation $Y$ of $\Hc$ acts on $L$ by $$ \lgl Y (\d) , x \rgl =
\lgl \d , Y(x) \rgl \, , \leqno (11) $$ and defines a derivation
of the Lie algebra $L$ where we recall that the Lie bracket in $L$
is given by, $$ \lgl [\d_1 , \d_2] , x \rgl = \lgl \d_1 \ot \d_2 -
\d_2 \ot \d_1 \, , \, \D x \rgl \qquad \fl \, x \in \Hc \, .
\leqno (12) $$ Let us now consider a map, $$ \ve \in \Cb
\backslash \{ 0 \} \ra \vp_{\ve} \in G \leqno (13) $$ such that
for any $x \in \Hc$, $\ov e (x) = 0$ one has, $$ \ve \ra \lgl
\vp_{\ve} , x \rgl \ \hbox{is a polynomial in} \ \frac{1}{\ve} \
\hbox{without constant term.} \leqno (14) $$ Thus $\ve \ra
\vp_{\ve}$ extends to a map from $P_1 (\Cb) \backslash \{ 0 \}$ to
$G$, such that $$ \vp_{\ify} = 1 \, . \leqno (15) $$ For such a
map we define its {\it residue} as the derivative at $\ify$, i.e.~as,
$$ {\rm Res} \, \vp = \lim_{\ve \ra \ify} \ve (\vp_{\ve} - 1)
\, . \leqno (16) $$ By construction ${\rm Res} \, \vp \in L$ is a
derivation $\Hc \ra \Cb$. When evaluated on $x \in \Hc$, ${\rm
Res} \, \vp$ is just the residue at $\ve = 0$ of the function $\ve
\ra \lgl \vp_{\ve} , x \rgl$.

We shall now assume that for any $t \in \Rb$ the following limit
exists for any $x \in \Hc$, $$ \lim_{\ve \ra 0} \ \lgl
\vp_{\ve}^{-1} \, \t_{t\ve} (\vp_{\ve}) , x \rgl \, . \leqno (17)
$$ Using (10), (4) and (5) we have, $$ \lgl \vp_{\ve}^{-1} \,
\t_{t\ve} (\vp_{\ve}) , x \rgl = \lgl \vp_{\ve} \ot \vp_{\ve} , (S
\ot \t_{t\ve}) \, \D x \rgl \, , \leqno (18) $$ so that with $\D x
= \sum x_{(1)} \, \ot \, x_{(2)}$ we get a sum of terms $\lgl
\vp_{\ve} , S \, x_{(1)} \rgl \, \lgl \vp_{\ve} , \t_{t\ve} (
x_{(2)}) \rgl$ $= P_1 \left( \frac{1}{\ve} \right) \, e^{kt\ve} \,
P_2 \left( \frac{1}{\ve} \right)$. Thus (17) just means that the
sum of these terms is holomorphic at $\ve = 0$. It is clear that
the value at $\ve = 0$ is then a polynomial in $t$, $$ \lgl F_t ,
x \rgl = \lim_{\ve \ra 0} \ \lgl \vp_{\ve}^{-1} \, \t_{t\ve}
(\vp_{\ve}) , x \rgl \, . \leqno (19) $$ Let us check that $t \ra
F_t \in G$ is a one parameter group, $$ F_{t_1 + t_2} = F_{t_1} \,
F_{t_2} \qquad \fl \, t_i \in \Rb \, . \leqno (20) $$ The group
$G$ is a topological group for the topology of simple convergence,
i.e., $$ \vp_n \ra \vp \quad \hbox{iff} \quad \lgl \vp_n , x \rgl
\ra \lgl \vp , x \rgl \qquad \fl \, x \in \Hc \, . \leqno (21) $$
Moreover, using (10) one checks that $$ \t_{t_1 \ve}
(\vp_{\ve}^{-1} \, \t_{t_2 \ve} (\vp_{\ve})) \ra F_{t_2} \qquad
\hbox{when} \ \ve \ra 0 \, . \leqno (22) $$ We then have $F_{t_1 +
t_2}$ $=$ ${\displaystyle \lim_{\ve \ra 0}} \ \vp_{\ve}^{-1} \,
\t_{(t_1 + t_2) \ve} \, (\vp_{\ve})$ $=$ ${\displaystyle \lim_{\ve
\ra 0}} \ \vp_{\ve}^{-1} \, \t_{t_1 \ve} (\vp_{\ve}) \, \t_{t_1
\ve} (\vp_{\ve}^{-1} \, \t_{t_2 \ve} (\vp_{\ve}))$ $=$ $F_{t_1} \,
F_{t_2}$.

This proves (20) and we let, $$ \b = \left(
\frac{\partial}{\partial t} \, F_t \right)_{t=0} \leqno (23) $$
which defines an element of $L$ such that, $$ F_t = \exp (t\b)
\qquad \fl \, t \in \Rb \, . \leqno (24) $$ As above, we view
$\Hc^*$ as an algebra on which $Y$ acts as a derivation by (11).
Let us prove,

\medskip

\noindent {\bf Lemma 1.} {\it Let $\ve \ra \vp_{\ve} \in G$
satisfy $(17)$ with $\vp_{\ve} = 1 + {\displaystyle
\sum_{n=1}^{\ify}} \ \frac{d_n}{\ve^n}$, $d_n \in \Hc^*$. One then
has $$  \ Y
(d_1) = \b \qquad Y \, d_{n+1} = d_n \, \b \qquad \fl \, n \geq 1  . $$ }

\medskip

\noindent {\it Proof.} Let $x \in \Hc$ and let us show that $$
\lgl \b , x \rgl = \lim_{\ve \ra 0} \, \ve \lgl \vp_{\ve} \ot
\vp_{\ve} , (S \ot Y) \, \D (x) \rgl \, . \leqno (25) $$

Using (18) we know by hypothesis that, $$ \lgl \vp_{\ve} \ot
\vp_{\ve} , (S \ot \t_{t\ve}) \, \D (x) \rgl \ra \lgl F_t , x \rgl
\leqno (26) $$ where the convergence holds in the space of
holomorphic functions of $t$ in say $\vert t \vert \leq 1$ so that
the derivatives of both sides at $t = 0$ are also convergent thus
yielding (25).

Now the function $\ve \ra \ve \, \lgl \vp_{\ve} \ot \vp_{\ve} , (S
\ot Y) \, \D x \rgl$ is holomorphic for $\ve \in \Cb \backslash \{
0 \}$ and also at $\ve = \ify \in P_1 (\Cb)$ since $\vp_{\ify} =
1$. Moreover by (25) it is also holomorphic at $\ve = 0$ and is
thus a constant, which gives, $$ \lgl \vp_{\ve} \ot \vp_{\ve} , (S
\ot Y) \, \D (x) \rgl = \frac{1}{\ve} \ \lgl \b , x \rgl \, .
\leqno (27) $$ Using the product in $\Hc^*$ this means that $$
\vp_{\ve}^{-1} \, Y (\vp_{\ve}) = \frac{1}{\ve} \, \b \, , \leqno
(28) $$ and multiplying by $\vp_{\ve}$ on the left, that, $$ Y
(\vp_{\ve}) = \frac{1}{\ve} \ \vp_{\ve} \, \b \, . \leqno (29) $$
One has $Y (\vp_{\ve}) = {\displaystyle \sum_{n=1}^{\ify}} \
\frac{Y (d_n)}{\ve^n}$ and $\frac{1}{\ve} \, \vp_{\ve} \, \b =
\frac{1}{\ve} \, \b + {\displaystyle \sum_{n=1}^{\ify}} \
\frac{1}{\ve^{n+1}} \, d_n \, \b$. Thus (29) gives the lemma.~\xx

\bigskip

In particular we get $Y (d_1) = \b$ and since $d_1$ is the
residue, ${\rm Res} \, \vp$, this gives, $$ \b = Y \, ({\rm Res}
\, \vp) \, , \leqno (30) $$ which shows that $\b$ is uniquely
determined by the residue of $\vp_{\ve}$.

We shall now write a formula for $\vp_{\ve}$ in terms of $\b$.
This is made possible by Lemma~1  which shows that $\b$ uniquely
determines $\vp_{\ve}$. What is not transparent from Lemma~1 is
that for $\b \in L$ the elements $\vp_{\ve} \in \Hc^*$ are
group-like, so that $\vp_{\ve} \in G$. In order to obtain a nice
formula we take the semi direct product of $G$ by $\Rb$ acting on
$G$ by the grading $\t_t$, $$ \wt G = G \, \semi_{\t} \, \Rb \, ,
\leqno (31) $$ and similarly we let $\wt L$ be the Lie algebra $$
\wt L = L \op \Cb \, Z_0 \leqno (32) $$ where the Lie bracket is
given by $$ [Z_0 , \a ] = Y (\a) \qquad \fl \, \a \in L \leqno
(33) $$ and extends the Lie bracket of $L$.

We view $\wt L$ as the Lie algebra of $\wt G$ in a way which will
become clear in the proof of the following,

\medskip

\noindent {\bf Theorem 2.} {\it Let $\ve \ra \vp_{\ve} \in G$
satisfy $(17)$ as above. Then with $\b = Y ({\rm Res} \, \vp)$ one
has, $$ \vp_{\ve} = \lim_{t \ra \ify} e^{-tZ_0} \, e^{t \left(
\frac{\b}{\ve} + Z_0 \right)} \, . $$ }

\medskip

The limit holds in the topology of simple convergence in $G$. Both
terms $e^{-t Z_0}$ and $e^{t \left( \frac{\b}{\ve} + Z_0 \right)}$
belong to $\wt G$ but their product belongs to $G$.

\medskip

\noindent {\it Proof.} We endow  $\Hc^*$ with the topology of
simple convergence on $\Hc$ and let $\t_t$ act by automorphisms of
the topological algebra $\Hc^*$ by (10). Let us first show, with,
$$ \vp_{\ve} = 1 + \sum_{n=1}^{\ify} \ \frac{d_n}{\ve^n} \, ,
\quad d_n \in \Hc^* \, , \leqno (34) $$ that the following holds,
$$ d_n = \int_{s_1 \geq s_2 \geq \cdots \geq s_n \geq 0} \t_{-s_1}
(\b) \, \t_{-s_2} (\b) \ldots \t_{-s_n} (\b) \, \Pi  \, ds_i \, .
\leqno (35) $$ For $n = 1$, this just means that, $$ d_1 =
\int_0^{\ify} \t_{-s} (\b) \, ds \, , \leqno (36) $$ which follows
from (30) and the equality $$ Y^{-1} (x) =  \int_0^{\ify} \t_{-s}
(x) \, ds \qquad \fl \, x \in \Hc \, , \ \ov e (x) = 0 \, . \leqno
(37) $$ We see from (37) that for $\a , \a' \in \Hc^*$ such that
$$ Y (\a) = \a' \, , \ \lgl \a , 1 \rgl = \lgl \a' , 1 \rgl = 0
\leqno (38) $$ one has, $$ \a = \int_0^{\ify} \t_{-s} (\a') \, ds
\, . \leqno (39) $$ Combining this equality with Lemma~1 and the
fact that $\t_s \in {\rm Aut} \, \Hc^*$ is an automorphism, gives
an inductive proof of (35). The meaning of this formula should be
clear, we pair both sides with $x \in \Hc$, and let, $$ \D^{(n-1)}
\, x = \sum x_{(1)} \ot x_{(2)} \ot \cdots \ot x_{(n)} \, . \leqno
(40) $$ Then the right hand side of (35) is just, $$ \int_{s_1
\geq \cdots \geq s_n \geq 0}  \, \lgl \b \ot \cdots \ot \b \ , \
\t_{-s_1} (x_{(1)}) \ot \t_{-s_2} (x_{(2)}) \cdots \ot \t_{-s_n}
(x_{(n)}) \rgl  \Pi  \, ds_i \leqno (41) $$ and the convergence of
the multiple integral is exponential since, $$ \lgl \b , \t_{-s}
(x_{(i)}) \rgl = O \, (e^{-s}) \qquad \hbox{for} \quad s \ra +
\ify \, . \leqno (42) $$ We see moreover that if $x$ is
homogeneous of degree $\deg (x)$, and if $n
> \deg (x)$, at least one of the $x_{(i)}$ has degree 0 so that $\lgl \b ,
\t_{-s} (x_{(i)}) \rgl = 0$ and (41) gives 0. This shows that the
pairing of $\vp_{\ve}$ with $x \in \Hc$ only involves finitely
many non zero terms in the formula, $$ \lgl \vp_{\ve} , x \rgl =
\ov e (x) + \sum_{n=1}^{\ify} \frac{1}{\ve^n} \ \lgl d_n , x \rgl
\, . \leqno (43) $$ With all convergence problems out of the way
we can now proceed to prove the formula of Theorem~2 without care
for convergence.

Let us first recall the expansional formula \cite{Ar}, $$
e^{(A+B)} = \sum_{n=0}^{\ify} \, \int_{\sum u_j = 1 , \, u_j \geq
0} \, e^{u_0 A} \, B e^{u_1 A} \ldots B e^{u_n A} \, \Pi  \, du_j
\leqno (44) $$ (cf.~\cite{Ar} for the exact range of validity of
(44)).

We apply this with $A = t Z_0$, $B = t \b$, $t > 0$ and get, $$
e^{t(\b + Z_0)} = \sum_{n=0}^{\ify} \ \int_{\sum v_j = t , \, v_j
\geq 0} \ e^{v_0 Z_0} \, \b e^{v_1 Z_0} \, \b \ldots \b e^{v_n
Z_0} \, \Pi  \, dv_j \, . \leqno (45) $$ Thus, with $s_1 = t -
v_0$, $s_1 - s_2 = v_1 , \ldots ,s_{n-1} - s_n = v_{n-1}$, $s_n =
v_n$ and replacing $\b$ by $\frac{1}{\ve} \, \b$, we obtain, $$
e^{t (\b / \ve + Z_0)} = \sum_{n=0}^{\ify} \ \frac{1}{\ve^n} \
\int_{t \geq s_1 \geq s_2 \geq \cdots \geq s_n \geq 0} \, e^{t
Z_0} \, \t_{-s_1} (\b) \ldots \t_{-s_n} (\b) \,  \Pi  \,ds_i .
\leqno (46) $$ Multiplying by $e^{-t Z_0}$ on the left and using
(41) thus gives, $$ \vp_{\ve} = \lim_{t \ra \ify} \ e^{-t Z_0} \,
e^{t (\b / \ve + Z_0)} \, . \leqno (47) $$ \hfill ~\xx

\bigskip

It is obvious conversely that this formula defines a family $\ve
\ra \vp_{\ve}$ of group-like elements of $\Hc^*$ associated to any
preassigned element $\b \in L$.

\medskip

\noindent {\bf Corollary 3.} {\it For any $\b \in L$ there exists
a (unique) map $\ve \ra \vp_{\ve} \in G$ satisfying $(17)$ and
$(34)$.}

\medskip

\section{The renormalization group flow}\label{secIII}
Let us now apply the above results to the group $G$ associated in
part~I to the Hopf algebra $\Hc$ of 1PI Feynman graphs of a
quantum field theory. We choose $\vp_6^3$ for simplicity. As
explained in part~I the group $G$ is a semi-direct product, $$ G =
G_0 \, \semi \, G_c \leqno (1) $$ of an abelian group $G_0$ by the
group $G_c$ associated to the Hopf subalgebra $\Hc_c$ constructed
on 1PI graphs with two or three external legs and fixed external
structure. Passing from $G_c$ to $G$ is a trivial step and we
shall thus concentrate on the group $G_c$. The unrenormalized
theory delivers, using dimensional regularization with the unit of
mass $\mu$, a loop, $$ \ve \ra \g_{\mu} (\ve) \in G_c \, , \leqno
(2) $$ and we first need to see the exact $\mu$ dependence of this
loop. We consider the grading of $\Hc_c$ and $G_c$ given by the
loop number of a graph, $$ L (\G) = I - V + 1 \leqno (3) $$ where
$I$ is the number of internal lines and $V$ the number of
vertices.

One has, $$ \g_{e^t \mu} (\ve) = \t_{t\ve} (\g_{\mu} (\ve)) \qquad
\fl \, t \in \Rb \, . \leqno (4) $$ Let us check this using the
formulas of section~3 of part~I. For $N=2$ external legs the
dimension $B$ of $\lgl \s , U_{\G} \rgl$ is equal to 0 by (12) of
loc.cit. Thus the $\mu$ dependence is given by $$
\mu^{\frac{\ve}{2} V_3} \leqno (5) $$ where $V_3$ is the number of
3-point vertices of $\G$. One checks that $\frac{1}{2} \, V_3 = L$
as required. Similarly if $N = 3$ the dimension $B$ of $\lgl \s ,
U_{\G} \rgl$ is equal to $\left( 1 - \frac{3}{2} \right) \, d +
3$, $d = 6 - \ve$ by (12) of loc.cit.~so that the $\mu$-dependence
is, $$ \mu^{\frac{\ve}{2} V_3} \, \mu^{-\ve / 2} \, . \leqno (6)
$$ But this time, $V_3 = 2L+1$ and we get $$ \mu^{\ve L}
\leqno (7) $$ as required.

We now reformulate a well known result, the fact that
counterterms, once appropriately normalized, are independent of
$m^2$ and $\mu^2$,

\medskip

\noindent {\bf Lemma 4.} {\it Let $\g_{\mu} = (\g_{\mu^-})^{-1} \,
(\g_{\mu^+})$ be the Birkhoff decomposition of $\g_{\mu}$. Then
$\g_{\mu^-}$ is independent of $\mu$.}

\medskip

As in part I we perform the Birkhoff decomposition with respect to
a small circle $C$ with center $D = 6$ and radius $<1$.

The proof of the lemma follows immediately from \cite{Col}. Indeed
the dependence in $m^2$ has in the minimal subtraction scheme the
same origin as the dependence in $p^2$ and we have chosen the
external structure of graphs (eq. (41) of part I) so that no $m^2$
dependence is left \footnote{This can be easily achieved by
maintaining non-vanishing  fixed external momenta.
$\gamma_{\mu^-}$ is independent on such external structures by
construction \cite{I}.}. But then, since $\mu^2$ is a dimensionful
parameter, it cannot be involved any longer.~\xx

\bigskip

\noindent {\bf Corollary 5.} {\it Let $\vp_{\ve} =
(\g_{\mu^-})^{-1} (\ve)$, then for any $t \in \Rb$ the following
limit exists in $G_c$, $$ \lim_{\ve \ra 0} \, \vp_{\ve}^{-1} \,
\t_{t\ve} (\vp_{\ve}) \, . $$ }

\medskip

In other words $\ve \ra \vp_{\ve} \in G_c$ fulfills the condition
(17) of section~2.

\medskip

\noindent {\it Proof.} The product $\vp_{\ve}^{-1} \, \g_{\mu}
(\ve)$ is holomorphic at $\ve = 0$ for any value of $\mu$. Thus by
(4), for any $t \in \Rb$, both $\vp_{\ve}^{-1} \, \g_{\mu} (\ve)$
and $\vp_{\ve}^{-1} \, \t_{\ve t} (\g_{\mu} (\ve))$ are
holomorphic at $\ve = 0$. The same holds for $\t_{-\ve t} \,
(\vp_{\ve}^{-1}) \, \g_{\mu} (\ve)$ and hence for the ratio $$
\vp_{\ve}^{-1} \, \g_{\mu} (\ve) \, (\t_{-\ve t} (\vp_{\ve}^{-1})
\, \g_{\mu} (\ve))^{-1} = \vp_{\ve}^{-1} \, \t_{-\ve t}
(\vp_{\ve}) \, . $$ \hfill ~\xx

\bigskip

We let $\g_- (\ve) = \vp_{\ve}^{-1}$ and translate the results of
section~2.

\medskip

\noindent {\bf Corollary 6.} {\it Let $F_t = {\displaystyle
\lim_{\ve \ra 0}} \, \g_- (\ve) \, \t_{t\ve} (\g_- \,
(\ve)^{-1})$. Then $F_t$ is a one parameter subgroup of $G_c$ and
$F_t = \exp (t\b)$ where $\b = Y \, {\rm Res} \, \vp_{\ve}$ is the
grading operator $Y$ applied to the residue of the loop $\g
(\ve)$.}

\medskip

In general, given a loop $\ve \ra \g (\ve) \in G$ it is natural to
define its {\it residue} at $\ve = 0$ by first performing the
Birkhoff decomposition on a small circle $C$ around $\ve = 0$ and
then taking, $$ {\rm Res}_{\ve = 0} \, \g =
\frac{\partial}{\partial \, u} \, (\vp_{1/u})_{u=0} \leqno (8) $$
where $\vp_{\ve} = \g_- (\ve)^{-1}$ and $\g (\ve) = \g_-
(\ve)^{-1} \, \g_+ (\ve)$ is the Birkhoff decomposition.

As shown in section~2, the residue or equivalently $\b = Y \, {\rm
Res}$ uniquely determines $\vp_{\ve} = \g_- (\ve)^{-1}$ and we thus
get, from Theorem~2,

\medskip

\noindent {\bf Corollary 7.} {\it The negative part $\g_- (\ve)$
of the Birkhoff decomposition of $\g_{\mu} (\ve)$ is independent
of $\mu$ and given by, $$ \g_- (\ve) = \lim_{t \ra \ify} \, e^{-t
\left( \frac{\b}{\ve} + Z_0 \right)} \, e^{t Z_0} \, . $$ }

\medskip

As above we adjoined the primitive element $Z_0$ to implement the
grading $Y$ (cf.~section~\ref{secII}). Our choice of the letter $\b$ is of
course not innocent and we shall see in section~5 the relation with the
$\b$-function.

\section{The action of $G_c$ on the coupling constants}\label{sec4}
We shall show in this section that the formula for the bare
coupling constant $g_0$ in terms of 1PI graphs, i.e. the
generating function, $$ g_0 = (g \, Z_1) \, (Z_3)^{-3/2} \leqno
(1) $$ where we consider the right hand side as a formal power
series with values in $\Hc_c$ given explicitly by, (with $\ell = L
(\G)$ the loop number of the graphs), $$ g_0=\left(
x+\sum_{\epsfysize=3mm\epsfbox{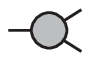}} x^{2l+1}
\frac{\Gamma}{S(\Gamma)} \right) \left(
1-\sum_{\epsfysize=3mm\epsfbox{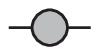}}  x^{2l}
\frac{\Gamma}{S(\Gamma)}\right)^{-3/2} \leqno (2) $$

 does define a
Hopf algebra homomorphism, $$ \Phi : \Hc_{\rm CM} \ra \Hc \leqno (3) $$
from the Hopf algebra $\Hc_{\rm CM}$ of coordinates on the group
of formal diffeomorphisms of $\Cb$ with $$ \vp (0) = 0 \, ,
\ \vp' (0) = 1 \, , \leqno (4) $$ to the Hopf algebra $\Hc_c$ of
1PI graphs.

This result is only valid if we perform on $\Hc_c$ the
simplification that pertains to the massless case, $m=0$, but
because of the $m$-independence of the counterterms all the
corollaries will be valid in general. The desired simplification
comes because in the case $m=0$ there is no need to indicate by a
cross left on an internal line the removal of a self energy
subgraph. Indeed and with the notations of part~I we can first of
all ignore all the ${\epsfysize=2.5mm\epsfbox{fig1b.eps}}_{(0)}$
since $m^2 = 0$, moreover the
$\raisebox{-1mm}{\epsfysize=4mm\epsfbox{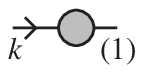}}$ yield a $k^2$
term which exactly cancels out with the additional propagator when
we remove the subgraph and replace it by $
\epsfysize=3mm\epsfbox{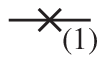}$. This shows that we can simply
ignore all these crosses and write coproducts in the simplest
possible way. To get familiar with this coproduct and with the
meaning of the Hopf algebra morphism (3) we urge the reader to
begin by the concrete computation done in the appendix, which
checks its validity up to order six in the coupling constant.

Let us now be more explicit on the meaning of formula (2). We
first expand $g_0$ as a power series in $x$ and get a series of
the form, $$ g_0 = x + \sum_2^{\ify} \a_{n} \, x^{n} \leqno
(5) $$ where the even coefficients $\a_{2n}$ are zero and the
coefficients $\a_{2n+1}$ are finite linear
combinations of products of graphs, so that, $$ \a_{2n+1} \in \Hc
\qquad \fl \, n \geq 1 \, . \leqno (6) $$ We let $\Hc_{\rm CM}$ be the Hopf algebra
 of the group of formal diffeomorphisms such that (4) holds. We take the
generators $a_n$ of $\Hc_{\rm CM}$ given by the equality $$ \vp (x) = x + \sum_{n \geq 2} a_n (\vp)
\, x^n \, . \leqno (7) $$ and define the coproduct in $\Hc_{\rm CM}$
by the equality $$   \lgl \D a_n  \, , \, \vp_1 \ot \vp_2   \rgl = a_n(\vp_2 \circ \vp_1)\leqno (8) $$
We then define uniquely the algebra homomorphism
$$ \Phi : \Hc_{\rm CM} \ra \Hc $$ by the condition,
$$ \Phi (a_n) =\a_{n}  \, . \leqno (9) $$  By construction
$\Phi$ is a morphism of algebras. We shall show that it is
comultiplicative, i.e. $$ (\Phi \ot \Phi) \, \D \, x = \D \, \Phi
(x) \qquad \fl \, x \in \Hc_{\rm CM} \leqno (10) $$ and comes from
a group morphism, $$ \rho : G_c \ra G_2 \, . \leqno (11) $$ where $G_2$ is the group
of characters of $\Hc_{\rm CM}$ which is by construction the opposite of the group
of formal diffeomorphisms. In
fact we shall first describe the corresponding Lie algebra
morphism, $ \rho$.

Let us first recall from part I that a 1PI graph $\G$ defines a
primitive element $(\G)$
 of $\Hc_{\rm gr}^*$ which only pairs
nontrivially with the monomial $\G$ of $\Hc$ and satisfies $\lgl (\G) \, ,\, \G \rgl =1$.
 We take the
following natural basis $\underline \G =S(\G) (\G)$ for the Lie
algebra of primitive elements of $\Hc_{\rm gr}^*$, labelled by 1PI
graphs with two or three external legs. By part I, theorem 2, their Lie
bracket is given by, $$ [\underline\G , \underline\G'] = \sum_{v'}
\underline{\G' \circ^{v'} \G} - \sum_v \underline{\G \circ^v \G' }
\, , \leqno (12) $$ where $\G' \circ_{v'} \G $ is the graph
obtained by grafting $ \G $ at the vertex $v'$ of $\G' $. (Our
basis differs from the one used in loc.~cit.~by an overall - sign,
but the present choice will be more convenient). In our context of
the simplified Hopf algebra the places where a given graph $\G$
can be inserted in another graph $\G'$ are no longer always
labelled by vertices of $\G'$. They are when $\G$ is a vertex
graph but when $\G$ is a self energy graph such places are just
labelled by the internal lines of $\G'$, as we could discard the
use of external structures and two-point vertices for self energy
graphs.

We also let $Z'_n$ be the natural basis of the Lie algebra of
primitive elements of $\Hc_{\rm CM}^*$ which corresponds to
 the vector fields $ x^{n+1} \ \frac{\partial}{\partial x}$.
 More precisely, $Z'_n$ is given as the linear form on $\Hc_{\rm CM}$ which only pairs
with the monomial $a_{n+1}$, $$\lgl Z'_n \, , \, a_{n+1} \rgl \, = 1 ,\leqno (13) $$
and the Lie bracket is given by, $$ [Z'_n ,
Z'_m] = (m-n) \, Z'_{n+m} \, . \leqno (14) $$ We then first prove,

\medskip

\noindent {\bf Lemma 8.} {\it  Let $\rho_{\G} = \frac{3}{2}$ for
$2$-point graphs and $\rho_{\G}=1$ for $3$-point graphs. The
equality $\rho (\underline\G) = \rho_{\G} \, Z'_{2\ell}$, where
$\ell = L (\G)$ is the loop number, defines a Lie algebra
homomorphism.}

\medskip

\noindent {\it Proof.} We just need to show that $\rho$ preserves
the Lie bracket. Let us first assume that $\G_1 , \G_2$ are vertex
graphs and let $V_i$ be the vertex number of $\G_i$. One has, $$ V
= 2L + 1 \leqno (15) $$ for any vertex graph $\G$. Thus the Lie
bracket $\rho \, [\underline\G_1 , \underline\G_2]$ provides $V_2
- V_1 = 2 (L_2 - L_1)$ vertex graph contributions all equal to
$Z'_{2 (L_1 + L_2)}$ so that $$ \rho \, ([\underline\G_1 ,
\underline\G_2]) = 2 (L_2 - L_1) \, Z'_{2 (L_1 + L_2)} \leqno (16)
$$ which is exactly $[\rho \, (\underline\G_1) , \rho \,
(\underline\G_2)]$ by (14).

Let then $\G_1$ and $\G_2$ be 2-point graphs. For any such graph
one has, $$ I = 3L - 1 \leqno (17) $$ where $I$ is the number of
internal lines of $\G$. Thus $\rho \, ([\underline\G_1 ,
\underline\G_2])$ gives $I_2 - I_1 = 3 (L_2 - L_1)$ 2-point graph
contributions, each equal to $\frac{3}{2} \ Z'_{2(L_1 + L_2)}$.
Thus, $$ \rho \, ([\underline\G_1 , \underline\G_2]) = \frac{3}{2}
\ 3 (L_2 - L_1) \, Z'_{2 (L_1 + L_2)} \leqno (18) $$ but the right
hand side is $\rho_{\G_1} \, \rho_{\G_2} \, 2 (L_2 - L_1) \, Z'_{2
(L_1 + L_2)}$ so that, $$ \rho \, ([\underline\G_1 ,\underline
\G_2]) = [\rho \, (\underline\G_1) , \rho \, (\underline\G_2)] \,
, \leqno (19) $$ as required.

Finally if say $\G_1$ is a 3-point graph and $\G_2$ a 2-point
graph, we get from $[\underline\G_1 , \underline\G_2]$ a set of
$V_2$ 2-point graphs minus $I_1$ 3-point graphs which gives, $$
\left( \frac{3}{2} \, V_2 - I_1 \right) \, Z'_{2 (L_1 + L_2)} \, .
\leqno (20) $$ One has $V_2 = 2 \, L_2$, $I_1 = 3 \, L_1$ so that
$\frac{3}{2} \, V_2 - I_1 = 3 (L_2 - L_1) = \rho_{\G_1} \,
\rho_{\G_2} \, 2 (L_2 - L_1)$ which gives (19) as required.~\xx

\bigskip

We now have the Lie algebra morphism $\rho$ and the algebra
morphism $\Phi$. To $\rho$ corresponds a morphism of groups, $$
\rho : G_c \ra G_2 \leqno (21) $$ and we just need to check that
the algebra morphism $\Phi$ is the transposed of $\rho$ on the
coordinate algebras, $$ \Phi (a) = a \circ \rho \qquad \fl \, a
\in \Hc_{\rm CM} \, . \leqno (22) $$ To prove (22) it is enough to
show that $\Phi$ is equivariant with respect to the action of the
Lie algebra $L$ of primitive elements of $\Hc_{\rm gr}^*$. More
precisely, given a primitive element $$ Z \in \Hc_{\rm gr}^* \ , \
\D Z = Z \ot 1 + 1 \ot Z \, , \leqno (23) $$ we let $\partial_Z$
be the derivation of the algebra $\Hc$ given by, $$
\partial_Z (y) = \lgl Z \ot {\rm id} , \D y \rgl \in \Hc \qquad \fl \, y
\in \Hc \, . \leqno (24) $$ What we need to check is the
following,

\medskip

\noindent {\bf Lemma 9.} {\it For any $a \in \Hc_{\rm CM}$, $Z \in
L$ one has $\partial_Z \, \Phi (a) = \Phi (\partial_{(\rho Z)}
(a))$.}

\medskip

\noindent {\it Proof.} It is enough to check the equality when $Z$
is of the form $\underline\G=S(\G) (\G)$ with the above notations.
Thus we let $\G$ be a 1PI graph and let $\partial_{\G}$ be the
corresponding derivation of $\Hc$ given by (24) with $Z = S(\G)
(\G)$. Now by definition of the primitive element $(\G)$ one has
(cf. (48) section 2 of part~I), $$ \lgl (\G) \ot {\rm id} , \D \G' \rgl = \sum n (\G
, \G'' ; \G') \, \G'' \leqno (25) $$ where the integer $n (\G ,
\G'' ; \G')$ is the number of subgraphs of $\G'$ which are
isomorphic to $\G$ while $\G' / \G \cong \G''$. By Theorem~2 of
part~I we have, $$ S (\G) \, S (\G'') \, n (\G , \G'' ; \G') = i
(\G , \G'' ; \G') \, S (\G') \leqno (26) $$ where $i (\G , \G'' ;
\G')$ is the number of times $\G'$ appears in $\G'' \circ \G$. We
thus get, $$
\partial_{\G} \, \frac{\G'}{S (\G')} = \sum i (\G , \G'' ; \G') \,
\frac{\G''}{S (\G'')} \leqno (27) $$ which shows that
$\partial_{\G}$ admits a very simple definition in the generators
$\frac{\G'}{S (\G')}$ of $\Hc$.

The derivation $\partial_{(\rho Z)}$ of $\Hc_{\rm CM}$ is also
very easy to compute. One has by construction, (Lemma~8), $$ \rho
(Z) = \rho_{\G} \, Z'_{2\ell} \qquad \ell = L (\G) \leqno (28) $$
and the derivation $d_k$ of $\Hc_{\rm CM}$ associated to the
primitive element $Z'_k$ of $\Hc_{\rm CM}^*$ is simply given, in
the basis $a_n \in \Hc_{\rm CM}$ by, $$ d_k (a_n) = (n-k) \,
a_{n-k} \, . \leqno (29) $$ We thus get, $$
\partial_{(\rho Z)} = \rho_{\G} \, d_{2\ell} \ , \quad \ell = L (\G) \, .
\leqno (30) $$ Now by construction both $\partial_Z \, \Phi$ and
$\Phi \, \circ \,
\partial_{(\rho Z)}$ are derivations from the algebra $\Hc_{\rm CM}$ to $\Hc$ viewed as a
 bimodule over $\Hc_{\rm CM}$, i.e. satisfy,
$$ \d (ab) = \d (a) \, \Phi (b) + \Phi (a) \, \d (b) \, . \leqno
(31) $$ Thus, to prove the lemma we just need to check the
equality, $$
\partial_{\G} \, \Phi (a_n) = \rho_{\G} \, \Phi (d_{2\ell} (a_n))
\qquad \ell = L(\G) \leqno (32) $$ or equivalently using the
generating function $$ g_0 = x + \sum \Phi (a_n) \, x^n \, ,
\leqno (33) $$ that $$
\partial_{\G} \, g_0 = \rho_{\G} \, x^{2\ell + 1} \,
\frac{\partial}{\partial x} \, g_0 \, . \leqno (34) $$ Now by
construction of $\Phi$ we have, $$ g_0 = (x \, Z_1) (Z_3)^{-3/2} \
, \ Z_3 = 1 - \d Z \leqno (35) $$ where, $$
Z_1=1+\sum_{\epsfysize=3mm\epsfbox{fig1a.eps}}
x^{2l}\frac{\Gamma}{S(\Gamma)},\;\delta Z=
\sum_{\epsfysize=3mm\epsfbox{fig1b.eps}}
x^{2l}\frac{\Gamma}{S(\Gamma)} \leqno (36) $$
Thus, since both $\partial_{\G}$ and $\frac{\partial}{\partial x}$
are derivations we can eliminate the denominators in (34) and
rewrite the desired equality, after multiplying both sides by
$(1-\d Z)^{5/2}$ as, $$ \ldisplaylinesno{ \left( x \,
\frac{\partial}{\partial \G} \, Z_1 \right) (1-\d Z) + \frac{3}{2}
\left( \frac{\partial}{\partial \G} \, \d Z \right) (x \, Z_1)
&(37)  \cr = \rho_\Gamma \left( x^{2\ell + 1} \left(
\frac{\partial}{\partial x} \, (x \, Z_1) \right) (1-\d Z) +
\frac{3}{2} ( x \, Z_1) \, x^{2\ell + 1} \,
\frac{\partial}{\partial x} \, \d Z \,\right) . \cr } $$ Both
sides of this formula are bilinear expressions in the 1PI graphs.
We first need to compute $\frac{\partial}{\partial \G} \, Z_1$ and
$\frac{\partial}{\partial \G} \, \d Z$. One has $$
\frac{\partial}{\partial\Gamma}Z_1=\sum_{\epsfysize=3mm
\epsfbox{fig1a.eps}}x^{2l+2l^\prime}c(\Gamma,\Gamma^\prime)
\frac{\Gamma^\prime}{S(\Gamma^\prime)}\leqno (38) $$
and, $$ \frac{\partial}{\partial\Gamma}\delta
Z=\sum_{\epsfysize=3mm
\epsfbox{fig1b.eps}}x^{2l+2l^\prime}c(\Gamma,\Gamma^\prime)
\frac{\Gamma^\prime}{S(\Gamma^\prime)},\leqno (39) $$
where $\ell = L(\G)$, $\ell' = L (\G')$ are the loop numbers and
the integral coefficient $c (\G , \G')$ is given by, $$ c(\G,\G')
= V' \ \hbox{if} \ \rho_{\G} = 1 \quad \hbox{and} \quad c(\G,\G')
= I' \ \hbox{if} \ \rho_{\G} = 3/2 \leqno (40) $$ (where $V'$ and
$I'$ are respectively the number of vertices and of internal lines
of $\G'$).

To prove (38) and (39) we use (27) and we get in both cases
expressions like (38), (39) with $$ c(\G,\G') = \sum_{\G''} \ i
(\G , \G' ; \G'') \, . \leqno (41) $$ But this is exactly the
number of ways we can insert $\G$ inside $\G'$ and is thus the
same as (40). Let now $\G_1$ be a 3-point graph and $\G_2$ a
2-point graph. The coefficient of the bilinear term, $$
\frac{\G_1}{S(\G_1)} \ \frac{\G_2}{S(\G_2)} \, , \leqno (42) $$ in
the left hand side of (37) is given by $$ \left( \frac{3}{2} \ c
(\G , \G_2) - c (\G , \G_1) \right) \, x^{2\ell + 2\ell_1 +
2\ell_2 + 1} \, . \leqno (43) $$ Its coefficient in the right hand
side of (37) is coming from the terms, $$ \ldisplaylinesno{
x^{2\ell + 1} \, \frac{\partial}{\partial x} \left( x^{2\ell_1 +
1} \, \frac{\G_1}{S(\G_1)} \right) \left( \frac{-\G_2}{S(\G_2)}
\right) \, x^{2\ell_2} \, + &(44)  \cr \frac{3}{2} \ x^{2\ell_1 +
1} \, \frac{\G_1}{S(\G_1)} \, x^{2\ell + 1} \,
\frac{\partial}{\partial x} \left( x^{2\ell_2} \,
\frac{\G_2}{S(\G_2)} \right) \cr } $$ which gives, $$ (3 \ell_2 -
2 \ell_1 - 1) \, x^{1+2\ell + 2\ell_1 + 2 \ell_2} \, . \leqno (45)
$$ We thus only need to check the equality, $$ \frac{3}{2} \ c (\G
, \G_2) - c (\G , \G_1) = (3 \ell_2 - 2\ell_1 - 1) \, \rho_{\G} \,
. \leqno (46) $$ Note that in general, for any graph with $N$
external legs we have, $$ V = 2 (L-1) + N \, , \ I = 3 (L-1) + N
\, . \leqno (47) $$ Let us first take for $\G$ a 3-point graph so
that $\rho_{\G} = 1$. Then the left hand side of (46) gives
$\frac{3}{2} \, V_2 - V_1 = \frac{3}{2} \, (2\ell_2) - (2 (\ell_1
- 1) + 3) = 3\ell_2 - 2\ell_1 - 1$.

Let then $\G$ be a 2-point graph, i.e. $\rho_{\G} = \frac{3}{2}$.
Then the left hand side of (46) gives $\frac{3}{2} \, I_2 - I_1 =
\frac{3}{2} \, (3\ell_2 - 1) - 3\ell_1 = \rho_{\G} \, (3 \ell_2 -
2\ell_1 - 1)$ which gives the desired equality.

Finally we also need to check the scalar terms and the terms
linear in $\G_1$ or in $\G_2$. The only scalar terms in the left
hand side of (37) are coming from $x \, \frac{\partial}{\partial
\G} \, Z_1 + \frac{3}{2} \, x \, \frac{\partial}{\partial \G} \,
\d Z$ and this gives, $$ x^{2\ell + 1} \, \rho_{\G} \, . \leqno
(48) $$ The only scalar term in the right hand side of (37) comes
from $x^{2\ell + 1}$ thus they fulfill (37).

The terms linear in $\G_1$ in the left hand side of (37) come only
from $x \, \frac{\partial}{\partial \G} \, Z_1$ if $\G$ is a
3-point graph and the coefficient of $\G_1 / S (\G_1)$ is thus, $$
c (\G , \G_1) \, x^{1 + 2\ell_1 + 2\ell} \, . \leqno (49) $$ In
the right hand side of (37) we just get $$ (2\ell_1 + 1) \, x^{1 +
2\ell_1 + 2\ell} \, . \leqno (50) $$ We thus need to check that $c
(\G , \G_1) = 2\ell_1 + 1$ which follows from (40) and (47) since
$V_1 = 2\ell_1 + 1$.

Similarly, if $\G$ is a 2-point graph the left side of (37) only
contributes by $x \, \frac{\partial}{\partial \G} \, Z_1 +
\frac{3}{2} \, x^{2\ell + 1} \, Z_1$, so that the coefficient of
$\G_1 / S(\G_1)$ is $$ \left(c (\G , \G_1) + \frac{3}{2} \right)
\, x^{1+2\ell + 2\ell_1} \, . \leqno (51) $$ In the right hand
side of (37) we get just as above $$ (2\ell_1 + 1) \, x^{1+2\ell +
2\ell_1} \leqno (52) $$ multiplied by $\rho_{\G} = 3/2$.

Now here, since $\G$ is a 2-point graph, we have $c (\G , \G_1) =
I_1 = 3 (\ell_1 - 1) + 3 = 3 \ell_1$ so that $$ c (\G , \G_1) +
\frac{3}{2} = \frac{3}{2} \, (2\ell_1 + 1) = \rho_{\G} (2\ell_1 +
1) $$ as required.

The check for terms linear in $\G_2$ is similar.~\xx

\bigskip

We can now state the main result of this section,

\medskip

\noindent {\bf Theorem 10.} {\it The map $\Phi = \Hc_{\rm CM} \ra
\Hc$ given by the effective coupling is a Hopf algebra
homomorphism. The transposed Lie group morphism is $\rho : G_c \ra
G_2$.}

\medskip

The proof follows from Lemma 9 which shows that the map from $G_c$ to $G_2$
given by the transpose of the algebra morphism $\Phi$ is the Lie
group morphism $\rho$. By construction the morphism
$\Phi$ is compatible with the grading $\Theta$ of ${\cal H}$
and $\alpha$ of ${\cal H}_{CM}$ given by ${\rm deg}(a_n)=n-1$
(cf.~\cite{CM}), one has indeed,
$$ \Phi\circ\alpha_t=\Theta_{2t}\circ\Phi,\;\forall t\in \Rb. \leqno (51)
$$
Finally we remark that our proof of Theorem 10
is similar to the proof of the equality $$
F_{\phi_1\phi_2}=F_{\phi_2}\circ F_{\phi_1}\leqno (52)$$
for the Butcher series used in the numerical integration of differential equations, but that the presence
of the $Z_3$ factor makes it much more involved in our case.

\section{The $\beta$-function and the
Birkhoff decomposition of the unrenormalized effective
coupling in the diffeomorphism group}\label{sec5}

Let us first recall our notations from part~I concerning the
effective action. We work in the Euclidean signature of space time
and in order to minimize the number of minus signs we write the
functional integrals in the form, $$ {\cal N} \int e^{S(\vp)} \, P
(\vp) \, [D \vp] \leqno (1) $$ so that the Euclidean action
is\footnote{We know of course that the usual sign convention is
better to display the positivity of the action functional.}, $$ S
(\vp) = - \frac{1}{2} \, (\partial_{\mu} \, \vp)^2 - \frac{1}{2}
\, m^2 \, \vp^2 + \frac{g}{6} \, \vp^3 \, . \leqno (2) $$ The
effective action, which when used at tree level in (1) gives the
same answer as the full computation using (2), is then given in
dimension $d = 6 - \ve$ by, $$ \ldisplaylinesno{ S_{\rm eff} (\vp)
= S (\vp) + \sum_{{\rm 1PI}} \, \frac{(\mu^{\ve / 2} \,
g)^{n-2}}{n!}  \int \, \frac{\G}{S(\G)} \ (p_1 , \ldots , p_n)
&(3)\cr \times \vp (p_1) \ldots \vp (p_n) \, \Pi \, dp_i \cr } $$
where, as in part~I, we do not consider tree graphs as 1PI and the
integral is performed on the hyperplane $\sum p_i = 0$. To be more
precise one should view the right hand side of (3) as a formal
power series  with values in the Hopf algebra $\Hc$. The theory
provides us with a loop $$ \g_{\mu} (\ve) = \g_- (\ve)^{-1} \,
\g_{\mu_+} (\ve) \leqno (4) $$ of characters of $\Hc$. When we
evaluate $\g_{\mu} (\ve)$ (resp.~$\g_-$, $\g_{\mu^+}$) on the
right hand side of (3) we get respectively the unrenormalized
effective action, the bare action and the renormalized effective
action (in the MS scheme).

Our notation is hiding the $g$-dependence of $\g_{\mu} (\ve)$, but this
dependence
 is entirely governed by the grading. Indeed with $t= \log(g)$
one has, with obvious notations,
$$
\g_{\mu,g} (\ve) = \Theta_{2t}(\g_{\mu,1} (\ve)) \leqno (5)
$$
Since $\Theta_{t}$ is an automorphism the same equality holds for both $\g_{\mu_+}$ and $\g_{-}$.

As in section 4 we restrict ourselves
to the massless case  and
let $\g_{\mu} (\ve) = \g_- (\ve)^{-1} \, \g_{\mu_+}(\ve)$
 be the Birkhoff decomposition of $\g_{\mu} (\ve)=\g_{\mu,1} (\ve)$.
\medskip

\noindent {\bf Lemma 11.} {\it Let $\rho : G_c \ra G_2$ be given by Theorem~10. Then
$\rho(\g_{\mu} (\ve))(g)$ is the unrenormalized  effective coupling constant,
$\rho(\g_{\mu_+} (0))(g)$ is the renormalized  effective coupling constant and
$\rho(\g_{-} (\ve))(g)$ is the bare coupling constant $g_0$.}

\medskip

This follows from (3) and Theorem~10. It is now straightforward to translate the results
 of the previous sections in terms of diffeomorphisms. The only subtle point to remember is that
the group $G_2$ is the opposite of the group of diffeomorphisms so that if we view $\rho$ as a map
 to diffeomorphisms it is an antihomomorphism,
$$ \rho(\g_1 \, \g_2)=\, \rho(\g_2) \circ \rho(\g_1)\, . \leqno
(6) $$
\medskip

\noindent {\bf Theorem 12.} {\it The renormalization group flow is the
image $\rho (F_{t})$ by $\rho : G_c \ra {\rm Diff}$
of the one parameter group $F_{t} \in
G_c$.}
\medskip

\noindent {\it Proof.} The {\it bare coupling constant}  $g_0$ governs the bare
action,
$$
 S_{\rm bare} (\vp_0) = -\frac{1}{2} \, (\partial_{\mu} \, \vp_0)^2 + \mu^{\ve /2} \
\frac{g_0}{6} \, \vp_0^3 \leqno (7)
$$
in terms of the bare field $\vp_0$.

Now when we replace $\mu$ by
$$
\mu' = e^t \, \mu \leqno (8)
$$
we can keep the bare action, and hence the physical theory, unchanged
provided we replace the
renormalized coupling constant $g$ by $g'$ where
$$
g_0 (\ve , g') = e^{-\ve \, \frac{t}{2}} \, g_0 (\ve , g) \, . \leqno
(9)
$$
By construction we have,
$$
g' = \psi_{\ve}^{-1} (e^{-\ve \frac{t}{2}} \, \psi_{\ve}(g)) \leqno (10)
$$
where $\psi_{\ve}$ is the formal diffeomorphism given by
 $$\psi_{\ve} = \,\rho \, (\g_- (\ve)) \, . \leqno (11)$$
 Now the behaviour for
$\ve \ra 0$ of $g'$ given by (10) is the same as for,
$$
\psi_{\ve}^{-1} \, \a_{\ve t/2} (\psi_{\ve}) \leqno (12)
$$
where $\a_s$ is the grading of ${\rm Diff}$ given as above by
$$
\a_s (\psi)(x) = e^{-s }\, \psi (e^{s}x) \, . \leqno (13)
$$
Thus, since the map $\rho$ preserves the grading,
$$
\rho \, (\t_t (\g)) = \a_{t/2} \, \rho (\g) \leqno (14)
$$
(by (51) of section~4), we see by Corollary~6 of section~3 that,
$$
g' \ra \rho (F_{t})g \quad \hbox{when} \ \ve \ra 0 \, . \leqno (15)
$$ ~\xx

\medskip

As a corollary we get of course,

\medskip

\noindent {\bf Corollary 13.} {\it The image by $\rho$ of $\b \in L$ is
the $\b$-function of the theory.}

\medskip

In fact all the results of section~3 now translate to the group $G_2$.
We get the formula for the bare coupling constant in terms of
the $\b$-function, namely,
$$
\psi_{\ve} = \lim_{t \ra \ify} e^{-t Z_0} \, e^{t \left( \frac{\b}{\ve}
+ Z_0 \right)} \leqno (16)
$$
where $Z_0 = x \ \frac{\partial}{\partial x}$ is the generator of
scaling. But we can also express the main result of part~I independently
of the group $G$ or of its Hopf algebra $\Hc$. Indeed the group
homomorphism $\rho : G \ra G_2$ maps the Birkhoff decomposition of
$\g_{\mu} (\ve)$ to the Birkhoff decomposition of $\rho (\g_{\mu} (\ve))$.
But we saw above that $\rho \, (\g_{\mu} (\ve))$ is just the {\it
unrenormalized} effective coupling constant. We can thus state,

\medskip

\noindent {\bf Theorem 14.} {\it Let the unrenormalized effective
coupling constant $g_{\rm eff} (\ve)$  viewed as a formal power series
in $g$ be considered as a loop of formal
diffeomorphisms
and let $g_{\rm eff} (\ve)= (g_{{\rm eff}_-})^{-1}(\ve) \, g_{{\rm eff}_+}(\ve)$
be its Birkhoff decomposition in the group  of formal
diffeomorphisms. Then the loop $g_{{\rm eff}_-} (\ve)$ is the bare
coupling constant and $g_{{\rm eff}_+} (0)$ is the renormalized
effective coupling.}

\medskip

Note that $G_2$ is naturally isomorphic to the opposite group of Diff so we
used the opposite order in the Birkhoff decomposition.

This result is very striking since it no longer involves the Hopf algebra
$\Hc$ or the group $G$ but only the idea of thinking of the effective
coupling constant as a formal diffeomorphism. The proof is immediate, by
combining Lemma~11, Theorem~10 of section 4 with Theorem~4 of part I.

Now in the same way as the Riemann-Hilbert problem and the Birkhoff
decomposition for the group $G = {\rm GL} (n,\Cb)$ are intimately related
to the classification of holomorphic $n$-dimensional vector bundles on $P_1
(\Cb) = C_+ \cup C_-$, the Birkhoff decomposition for the group $G_2 = {\rm
Diff}^0$ is related to the classification of one dimensional complex (non
linear) bundles,
$$
P = (C_+ \ts X) \cup_{g_{\rm eff}} (C_- \ts X) \, . \leqno (17)
$$
Here $X$ stands for a formal one dimensional fiber and $C_{\pm}$ are, as in
part I, the components of the complement in $P_1 (\Cb)$ of a small circle
around $D$. The total space $P$ should be thought of as a 2-dimensional
complex manifold which blends together the $\ve = D-d$ and the coupling
constant of the theory.

\section{Conclusions}\label{sec6}
We showed in this paper that the group $G$
of characters of the Hopf algebra ${\cal H}$ of Feynman graphs
plays a key role in the geometric understanding of the basic ideas of
renormalization
including the renormalization group and the $\beta$-function.

We showed in particular that the group $G$
acts naturally on the complex space
$X$ of dimensionless coupling constants of the theory.
Thus, elements of $G$ are a refined form
of diffeomorphisms of $X$ and as such should be called
diffeographisms.

The action of these diffeographisms on the space of coupling
constants allowed us first of all to read off directly
the bare coupling constant and
the renormalized one from the Riemann--Hilbert
decomposition of the unrenormalized effective coupling constant
viewed as a loop of formal
diffeomorphisms. This showed that renormalization is intimately
related
with the theory of non-linear complex bundles on the Riemann sphere
of the dimensional regularization parameter $\ve$.
It also allowed us to lift
both the renormalization group and the $\beta$-function
as the asymptotic scaling in the group of diffeographisms.
This used the full power of the Riemann--Hilbert decomposition
together
with the invariance of $\gamma_-(\ve)$ under a change of unit
of mass. This gave us a completely streamlined proof of the existence
of the renormalization group and more importantly a closed
formula of scattering nature,
delivering the full higher pole structure of minimal
subtracted counterterms in terms of the residue.

In the light of the predominant role of the residue in NCG
we expect this type of formula to help us to decipher the message
on space-time geometry burried in the need for renormalization.

Moreover, thanks to \cite{K3} the previous results no longer
depend upon dimensional regularization but can be formulated in
any regularization or renormalization scheme. Also, we could
discard a detailed discussion of anomalous dimensions, since it is
an easy corollary \cite{BK} of the knowledge of the
$\beta$-function.

For reasons of simplicity our analysis was limited to the case of
one coupling constant. The generalization to a higher dimensional
space $X$ of coupling constants is expected to involve the same
ingredients as those which appear in higher dimensional
diffeomorphism groups and Gelfand-Fuchs cohomology \cite{CM}. We
left aside the detailed study of the Lie algebra of
diffeographisms and its many similarities with the Lie algebra of
formal vector fields. This, together with the interplay between
Hopf algebras, rational homotopy theory, BRST cohomology, rooted
trees and shuffle identities will be topics of future joint work.

\section{Appendix: up to three loops}\label{sec7}
We now want to check the Hopf algebra homomorphism ${\cal
H}_{CM}\to{\cal H}$ up to three loops as an example. We regard
$g_0$ as a series in a variable $x$ (which can be thought
of as a physical coupling) up to order $x^6$, making use of
$g_0=xZ_1Z_3^{-3/2}$ and the expression of the $Z$-factors in
terms of 1PI Feynman graph. The challenge is then to confirm that
the coordinates $\delta_n$ on $G_2$, implicitly defined by \cite{CM}
$$\log\left[g_0(x)^\prime\right]^{(n)}$$  commute with the
Hopf algebra homomorphism: calculating the coproduct $\Delta_{CM}$
of $\delta_n$ and expressing the result in Feynman graphs must
equal the application of the coproduct $\Delta$ applied to
$\delta_n$ expressed in Feynman graphs.

By (2) of section \ref{sec4} we write $g_0=xZ_1Z_3^{-3/2}$,
\[
Z_1=1+\sum_{k=1}^\infty z_{1,2k}x^{2k},
\]
\[
Z_3=1-\sum_{k=1}^\infty z_{3,2k}x^{2k},
\]
and
\[
Z_g=Z_1Z_3^{-3/2},\;z_{i,2k}\in {\cal H}_c,\;i=1,3,
\]
as formal series in $x^2$. Using
\[
\log\left(\frac{\partial}{\partial x}
xZ_g\right)=\sum_{k=1}^\infty\frac{\delta_{2k}}{(2k)!}x^{2k},
\]
which defines $\delta_{2k}$ as the previous generators $a_n(\phi)$
of coordinates of $G_2$, we find
\begin{eqnarray}
\frac{1}{2!}\delta_2 \equiv\tilde{\delta_2} & = &
3z_{1,2}+\frac{9}{2}z_{3,2},\label{e1}\\ \frac{1}{4!}\delta_4
\equiv\tilde{\delta_4} & = &
5[z_{1,4}+\frac{3}{2}z_{3,4}]-\frac{9}{2}z_{1,2}^2
-6z_{1,2}z_{3,2}-\frac{3}{4}z_{3,2}^2,\label{e2}\\
\frac{1}{6!}\delta_6 \equiv\tilde{\delta_6} & = &
9z_{1,2}^3+18z_{1,2}^2z_{3,2}-5[3z_{1,2}z_{1,4}+\frac{3}{2}z_{3,2}z_{3,4}]
\label{e3}\\ & & +12[ z_{1,2}z_{3,2}^2-z_{1,2}
z_{3,4}-z_{1,4}z_{3,2}]+7[z_{1,6}+\frac{1}{2}z_{3,2}^3
+\frac{3}{2}z_{3,6}].\nonumber
\end{eqnarray}
The algebra homomorphism ${\cal H}_c\to{\cal H}$
of section(\ref{sec4})
is effected by expressing the $z_{i,2k}$ in Feynman graphs,
with 1PI graphs
with three external legs contributing to $Z_1$, and 1PI graphs
with two external legs, self-energies, contributing to $Z_3$.

Explicitly, we have  {\large
\begin{eqnarray*} z_{1,2} & = & \v,\\ z_{3,2}
&= & \frac{1}{2}\p,\\ z_{1,4} &= &
\vlv+\vuv+\vdv+\frac{1}{2}\left[\vrp+\vup+\vdp\right]+\frac{1}{2}\w,\\
z_{3,4} &= & \frac{1}{2}\left[\pdp+\pv\right].\\
\end{eqnarray*}}
The symmetry factor {\large $$2=S\left(\w\right)$$} is most
obvious if we redraw $$\epsfysize=6mm\epsfbox{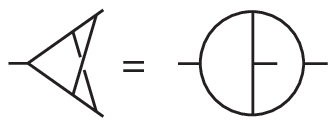}.$$
Further, we have
 {\large\begin{eqnarray*}
z_{1,6} & = &
\vlvlv+\vlvuv+\vlvdv+\vuvlv+\vuvuv+\vuvdv+\vdvlv+\vdvuv+\vdvdv\\
 & &
+\frac{1}{2}\left[\vlvup+\vlvdp+\vlvrp+\vuvup+\vuvdp+\vuvrp\right.\\
 & & \left.+\vdvup+\vdvdp+\vdvrp
\right] +\vluvv+\vldvv+\vudvv\\ & &
+\frac{1}{2}\left[\vlrvp+\vldvp+\vluvp+\vurvp+\vudvp+\vuuvp\right.\\
 & & \left.+\vdrvp+\vddvp+\vduvp\right]\\
 & &
+\frac{1}{4}\left[\vuupp+\vrrpp+\vddpp+\vurpp+\vdrpp+\vudpp\right]\\
 & &
+\frac{1}{2}\left[\wlv+\wurv+\wdrv\right]+\wulv\\
 & &
+\frac{1}{2}\left[\wulp+\wurp+\wdlp+\wdrp+\wmlp+\wmrp\right]\\
 & &
+\frac{1}{2}\left[\vlw+\vuw+\vdw\right]\\
 & &
+\frac{1}{2}\left[\vupv+\vrpv+\vdpv+\vupdp+\vrpdp+\vdpdp\right]+\mbox{\small
primitive terms},
\end{eqnarray*}}
and
 {\large\begin{eqnarray*}
z_{3,6} & = &
\frac{1}{2}\pdpdp+\frac{1}{8}\pdupp+\frac{1}{4}\pddpp
+\frac{1}{2}\pdpv+\frac{1}{2}\pw\nonumber\\
 & &
+\frac{1}{2}\plrvv+\frac{1}{4}
\pvmp+\frac{1}{2}\pvlp+\frac{1}{2}\pvrp+\pvuv.
\end{eqnarray*}}
Here, primitive terms refers to 1PI three-loop vertex graphs
without subdivergences. They fulfil all desired identities below
trivially, and are thus not explicitly given.
On the level of diffeomorphisms, we have the coproducts
\begin{eqnarray}
\Delta_{CM}[\delta_4] & = & \delta_4\ot1+1\ot\delta_4+4\delta_2\ot\delta_2,\\
\Delta_{CM}[\delta_6] & = & \delta_6\ot1+1\ot\delta_6+
20\delta_2\ot\delta_4+6\delta_4\ot\delta_2\nonumber\\
 & & +28\delta_2^2\ot\delta_2,\label{ed6}
\end{eqnarray}
where we skip odd gradings.

We have to check that the coproduct $\Delta$ of Feynman graphs
reproduces these results.

Applying $\Delta$ to the rhs of (\ref{e2}) gives, using the
expressions for $z_{i,k}$ in terms of Feynman graphs,
 {\large\begin{eqnarray*}
\Delta(\tilde{\delta_4}) & = &
6\v\ot\v+\frac{9}{2}\left[\v\ot\p+\p\ot\v\right]\\
 & & +\frac{27}{8}\p\ot\p+\tilde{\delta_4}\ot 1+1\ot
\tilde{ \delta_4}.
\end{eqnarray*}}
This has to be compared with $\tilde{\delta_4}\ot 1+1\ot \tilde{
\delta_4}+\frac{2!2!}{4!}4\tilde{\delta_2}\ot\tilde{\delta_2}$,
which matches perfectly, as
 {\large\begin{eqnarray*}
\tilde{\delta_2}\ot\tilde{\delta_2} & = & 9\v\ot\v+\frac{27}{4}
\left[\v\ot\p\;+\;\p\ot\v\right]\\ & & +\frac{81}{16}\p\ot\p.
\end{eqnarray*}}

After this warming up, let us do the check at order $g^6$, which
will be much more demanding, as the coproduct will be
noncocommutative now.

We need that $\Delta_{CM}(\delta_6)$ in (\ref{ed6}) is equivalent
to
\begin{eqnarray}
\Delta_{CM}(\tilde{\delta_6}) & = & \tilde{\delta_6}\ot
1+1\ot\tilde{\delta_6}+20\frac{2!4!}{6!}\tilde{\delta_2}
\ot\tilde{\delta_4}+6\frac{2!4!}{6!}\tilde{\delta_4}\ot\tilde{\delta_2}
\nonumber\\
 & & +28\frac{2!2!2!}{6!}\tilde{\delta_2}^2\ot\tilde{\delta_2}.
\end{eqnarray}
Applying the Hopf algebra homomorphism to Feynman graphs
on both sides of the tensor product delivers
 {\large\begin{eqnarray}
\Delta(\tilde{\delta_6}) & = & \tilde{\delta_6}\ot
1+1\ot\tilde{\delta_6}\nonumber\\
 & & +\frac{28}{90}\left[3\v+\frac{9}{4}\p\right]^2\ot\left[3\v+
\frac{9}{4}\p\right]\nonumber\\
 & &
 +\frac{6}{15}\left[5\left[\vlv+\vuv+\vdv+
 \frac{1}{2}\left[\vrp+\vup+\vdp+\w\right]\right.\right.\nonumber\\
  & &
\left.\left.
+\frac{3}{2}\left[\frac{1}{2}\pdp+\frac{1}{2}\pv\right]\right]\right.\nonumber\\
 & & \left. -3\left[\frac{3}{2}\v\v+\v\p+\frac{1}{16}\p\p\right]\right]\nonumber\\
 & & \ot \left[3\v+\frac{9}{4}\p\right]\nonumber\\
 & & +\frac{20}{15} \left[3\v+\frac{9}{4}\p\right]\nonumber\\
 & & \ot
5\left[\vlv+\vuv+\vdv+\frac{1}{2}\left[\vrp+\vup+\vdp+\w\right]
\right.\nonumber\\
 & & \left.
+\frac{3}{2}\left[\frac{1}{2}\pdp+\frac{1}{2}\pv\right]\right]
\nonumber\\
 & &  -3\left[\frac{3}{2}\v\v+\v\p+\frac{1}{16}\p\p\right].
\end{eqnarray}}
Multiplying this out, we find the following result
 {\large\begin{eqnarray}
\Delta(\tilde{\delta_6})  =  \tilde{\delta_6} & \ot & 1\nonumber\\
+1 & \ot & \tilde{\delta_6}\nonumber\\ +10\v & \ot & \w\nonumber\\
+3\w & \ot & \v\nonumber\\ +\frac{9}{4}\w & \ot &
\p\nonumber\\+\frac{15}{2}\p & \ot & \w\nonumber\\
 +\frac{9}{4}\left[\vrp+\vup+\vdp\right] & \ot & \p\nonumber\\
+\frac{15}{2}\p & \ot & \left[\vrp+\vup+\vdp\right]\nonumber\\
+\frac{9}{2}\left[\vlv+\vuv+\vdv\right] & \ot & \p\nonumber\\
+15\p & \ot & \left[\vlv+\vuv+\vdv\right]\nonumber\\
+\frac{9}{2}\pv & \ot & \v\nonumber\\+15\v & \ot & \pv\nonumber\\
+\frac{9}{2}\pdp & \ot & \v\nonumber\\+15\v & \ot & \pdp\nonumber\\
+\frac{27}{8}\pv & \ot & \p\nonumber\\
+\frac{45}{4}\p & \ot & \pv\nonumber\\
+\frac{27}{8}\pdp & \ot & \p\nonumber\\
+\frac{45}{4}\p & \ot & \pdp\nonumber\\
+3\left[\vrp+\vup+\vdp\right] & \ot & \v\nonumber\\
+10\v & \ot & \left[\vrp+\vup+\vdp\right]\nonumber\\
+6\left[\vlv+\vuv+\vdv\right] & \ot & \v\nonumber\\
+20\v & \ot & \left[\vlv+\vuv+\vdv\right]\nonumber\\
-9\p & \ot & \v\p\nonumber\\-\frac{9}{2}\p\p & \ot & \v\nonumber\\
-\frac{3}{4}\v & \ot & \p\p\nonumber\\
-\frac{9}{16}\p & \ot & \p\p\nonumber\\
-\frac{27}{2}\p & \ot & \v\v\nonumber\\-12\v & \ot & \v\p\nonumber\\
-18\v & \ot & \v\v\nonumber\\+\frac{27}{4}\p\v & \ot & \p\nonumber\\
+\p\p & \ot & \p\nonumber\\+9\p\v & \ot & \v\nonumber\\
+\frac{9}{4}\v\v & \ot & \p\nonumber\\+3\v\v & \ot & \v\label{e4}.
\end{eqnarray}}
Now we have to compare with $\Delta(\tilde{\delta_6})$,
so we first apply the homomorphism to graphs and use the coproduct
$\Delta$ on them. For this,
we need
 {\large\begin{eqnarray}
\Delta[z_{1,2}] & = & \v\ot 1+1\ot \v,\\ \Delta[z_{3,2}] & = &
\p\ot 1+1\ot \p,\\ \Delta[z_{1,4}] & = & z_{1,4}\ot 1+1\ot
z_{1,4}+3\v\ot\v+\frac{3}{2}\p\ot\v,\\ \Delta[z_{3,2}] & = &
z_{3,4}\ot 1+1\ot z_{3,4}+\frac{1}{2}\p\ot\p+\v\ot\p,\\
\Delta[z_{1,6}] & = & z_{1,6}\ot 1+1\ot
z_{1,6}+3\v\ot\left[\vlv+\vuv+\vdv\right] \\
 & &
+3\left[\vlv+\vuv+\vdv\right] \ot\v\nonumber\\
 & & +2\v\ot\left[\vlv+\vuv+\vdv\right]\nonumber\\
 & &
+3\v\v\ot\v+\p\ot\left[\vrp+\vup+\vdp\right]\nonumber\\
 & & +\frac{3}{2}\p\p\ot\v\nonumber\\
 & & +\frac{3}{2}\v\ot\left[\vrp+\vup+\vdp\right]+\frac{3}{2}\pv\ot\v\nonumber\\
 & & +\frac{1}{2}\p\ot\left[\vrp+\vup+\vdp\right]+\frac{3}{2}\pdp\ot\v\nonumber\\
& &
+\frac{3}{2}\p\ot\left[\vlv+\vuv+\vdv\right]\nonumber\\
 & & +\v\ot\left[\vrp+\vup+\vdp\right]
\nonumber\\
 & & +\frac{9}{2}\v\p\ot\v+\frac{3}{2}\p\ot\left[\vlv+\vuv+\vdv\right]\nonumber\\
 & & +\frac{3}{2}\left[\vrp+\vup+\vdp\right]\ot\v+\frac{3}{2}\w\ot\v\nonumber\\
 & & +\frac{5}{2}\v\ot\w+3\p\ot\w,\nonumber\\
\Delta[z_{3,6}] & = & z_{3,6}\ot 1+1\ot
z_{3,6}+\frac{1}{2}\p\ot\pdp\\
 & & +\frac{1}{2}\pdp\ot\p+\frac{1}{4}\p\ot\pdp\nonumber\\
 & & +\frac{1}{8}\p\p\ot\p
\nonumber\\
 & & +\frac{1}{2}\p\ot\pdp+\frac{1}{4}\p\p\ot\p+\v\ot\pdp\nonumber\\
 & & +\frac{1}{2}\pv\ot\p+\w\ot\p+\frac{1}{2}\v\v\ot\p\nonumber\\
 & & +\v\ot\pv+\frac{1}{4}\p\ot\pv\nonumber\\
  & & +\frac{1}{2}\vrp\ot\p+\p\ot\pv
\nonumber\\
 & & +\frac{1}{2}\left[\vup+\vdp\right]\ot\p+\v\ot\pdp+\v\ot\pv\nonumber\\
  & & +\left[\vlv+\vuv+\vdv\right]\ot\p.\nonumber
\end{eqnarray}}
It is now only a matter of using the rhs of
(\ref{e3}) for $\tilde{\delta_6}$
to confirm that we reproduce the result (\ref{e4}). For example,
for the contribution to $\v\ot\w$ in $\Delta(\tilde{\delta_6})$ we
find
 {\large\[
-\frac{5\times 3}{2}\v\ot\w+7\times \frac{5}{2}\v\ot\w=10\v\ot\w,
\]}
as desired. Similarly, one checks all of the 32 tensorproducts of
(\ref{e4}).

\section*{Acknowledgements}
Both authors thank the IHES for generous support during this
collaboration. D.K.~is grateful to the DFG for a Heisenberg
Fellowship.

\end{document}